\documentclass[floatfix,aps,prc,twocolumn,10pt,superscriptaddress,showpacs]{revtex4-2}

\usepackage{graphicx}  
\usepackage{amssymb}   
\usepackage{hyperref}  
\usepackage{lineno}
\usepackage{rotating}
\usepackage{float}
\usepackage{lineno}

\begin{document}

\title{Probing for high momentum protons in \texorpdfstring{$^4$He via the } \texorpdfstring{$^4He(e,e'p)X$} reactions}
	
	\newcommand*{\JLAB}{Thomas Jefferson National Accelerator Facility, Newport News, VA 23606}
	\newcommand*{\TLV}{Tel Aviv University, Tel Aviv 69978, Israel}
	\newcommand*{\MIT}{Massachusetts Institute of Technology, Cambridge, MA 02139}
	\newcommand*{\KENT}{Kent State University, Kent, OH 44242}
	\newcommand*{\DOMINION}{Old Dominion University, Norfolk, VA 23529}
	\newcommand*{\CALIF}{California State University, Los Angeles, Los Angeles, CA 90032}
	\newcommand*{\Hampton}{Hampton University, Hampton, VA 23668}
	\newcommand*{\PENNSYLVANIA}{Pennsylvania State University, State College, PA 16801}
	\newcommand*{\Paris}{Institut de Physique Nucl\'{e}aire (UMR 8608), CNRS/IN2P3 - Universit\'e Paris-Sud, F-91406 Orsay Cedex, France}
	\newcommand*{\Syracuse}{Syracuse University, Syracuse, NY 13244}
	\newcommand*{\Kentucky}{University of Kentucky, Lexington, KY 40506}
	\newcommand*{\William}{College of William and Mary, Williamsburg, VA 23187}
	\newcommand*{\Virginia}{University of Virginia, Charlottesville, VA 22904}
	\newcommand*{\Halifax}{Saint Mary's University, Halifax, Nova Scotia, Canada}
	\newcommand*{\Glasgow}{University of Glasgow, Glasgow G12 8QQ, Scotland, United Kingdom}
	\newcommand*{\Temple}{Temple University, Philadelphia, PA 19122}
	\newcommand*{\Argonne}{Physics Division, Argonne National Laboratory, Lemont, IL 60439}
	\newcommand*{\Berkeley}{Lawrence Berkeley National Laboratory, Berkeley, California 94720, USA}
	\newcommand*{\China}{China Institute of Atomic Energy, Beijing, China}
	\newcommand*{\NRCN}{Nuclear Research Center Negev, Beer-Sheva, Israel}
	\newcommand*{\Catania}{Universita di Catania, Catania, Italy}
	\newcommand*{\Duquesne}{Duquesne University, Pittsburgh, PA 15282}
	\newcommand*{\Pittsburgh}{Carnegie Mellon University, Pittsburgh, PA 15213}
	\newcommand*{\LongwoodUniv}{Longwood University, Farmville, VA 23909}
	\newcommand*{\Florida}{Florida International University, Miami, FL 33199}
	\newcommand*{\Tallahassee}{Florida State University, Tallahassee, FL 32306}
	\newcommand*{\INFN}{INFN, Sezione Sanit\`{a} and Istituto Superiore di Sanit\`{a}, 00161 Rome, Italy}
	\newcommand*{\INFNBari}{INFN, Sezione di Bari and University of Bari, I-70126 Bari, Italy}
	\newcommand*{\Ohio}{Ohio University, Athens, OH 45701}
	\newcommand*{\Tennessee}{University of Tennessee, Knoxville, TN 37996}
	\newcommand*{\Kharkov}{Kharkov Institute of Physics and Technology, Kharkov 61108, Ukraine}
	\newcommand*{\LOSALAMOS}{Los Alamos National Laboratory, Los Alamos, NM 87545}
	\newcommand*{\Duke}{Duke University, Durham, NC 27708}
	\newcommand*{\Texas}{University of Texas, Houston, TX 77030}
	\newcommand*{\Texasam} {Texas A\&M University,Kingsville, TX 78363} 
	\newcommand*{\Seoul}{Seoul National University, Seoul, Korea}
	\newcommand*{\Indiana}{Indiana University, Bloomington, IN 47405}
	\newcommand*{\Hampshire}{University of New Hampshire, Durham, NH 03824}
	\newcommand*{\Blacksburg}{Virginia Polytechnic Inst. and State Univ., Blacksburg, VA 24061}
	\newcommand*{\France}{Universit\'e Blaise Pascal/IN2P3, F-63177 Aubi\`ere, France}
	\newcommand*{\Mississippi}{Mississippi State University, Mississippi State, MS 39762}
	\newcommand*{\Austin}{The University of Texas at Austin, Austin, Texas 78712}
	\newcommand*{\Norfolk}{Norfolk State University, Norfolk, VA 23504}
	\newcommand*{\Lanzhou}{Lanzhou University, Lanzhou, China}
	\newcommand*{\Hebrew}{Racah Institute of Physics, Hebrew University of Jerusalem, Jerusalem, Israel}
	\newcommand*{\Rutgers}{Rutgers, The State University of New Jersey, Piscataway, NJ 08855}
	\newcommand*{\Yerevan}{Yerevan Physics Institute, Yerevan 375036, Armenia}
	\newcommand*{\Ljubljana}{Faculty of Mathematics and Physics, University of Ljubljana, Ljubljana, Slovenia}
	\newcommand*{\Michigan}{Northern Michigan University, Marquette, MI 49855}
	\newcommand*{\Hefei}{University of Science and Technology, Hefei, China}	
	\newcommand*{\Jozef}{Jozef Stefan Institute, Ljubljana, Slovenia}
	\newcommand*{\Ecole}{CEA Saclay, F-91191 Gif-sur-Yvette, France}
	\newcommand*{\Massachusetts}{University of Massachusetts, Amherst, MA 01006}
	\newcommand*{\Bulgaria}{Bulgarian Academy of Sciences, Bulgaria}
	\newcommand*{\Spain}{Computense University de Madrid, Spain}
	\newcommand*{\VirginiaUnion}{Virginia Union University, Richmond, VA 23220}
	\author{S. Iqbal}
	\affiliation{\CALIF}
	\author{F. Benmokhtar} 
\thanks{Contact person \email {benmokhtarf@duq.edu}}
	\affiliation{\Duquesne}
	\author{M. Ivanov}
	\affiliation{\Bulgaria}
	\author{N. See}
	\affiliation{\CALIF}
	\author{K. Aniol}
	\affiliation{\CALIF} 
	\author{D.~W. Higinbotham}
	\affiliation{\JLAB}
	\author{C. Boyd}
	\affiliation{\Duquesne}
		\author{A. Gadsby}
	\affiliation{\Duquesne}
    \author{J. S. Goodwill}
	\affiliation{\Duquesne}
	\author{D. Finton}
	\affiliation{\Duquesne}
% ...
	
\collaboration{The Jefferson Lab Hall A Collaboration}

	\date{\today}
	
\begin{abstract}	
Experimental cross sections for the $^4He(e,e'p)X$ reactions in the missing energy range from 0.017 to 0.022 GeV and up to a missing momentum of 0.632 GeV/$c$ at $x_B=1.24$ and $Q^2$=2(GeV/$c$)$^2$ are reported.  
The data are compared to Relativistic Distorted Wave Impulse Approximation(RDWIA) calculations for the $^4He(e,e'p)^3H$ channel.
Significantly more events are observed for $p_{m} \geq 0.45$ GeV/$c$ than are predicted by the theoretical model and a striking fluctuation in the ratio of data to the theoretical model around $p_{m} = 0.3~$GeV/$c$ are possible signals of initial-state multi-nucleon correlations.
\end{abstract}
	
	\pacs{13.60.Hb, 25.10.+s, 25.30.Fj}
   \keywords{High momentum protons, $^4He$}
		
	\maketitle

\section{Introduction}
Nucleon momentum distributions in atomic nuclei are known to be governed by an average nuclear potential plus additional nucleon-nucleon and nucleon-multi-body interactions \cite{kelly2002} \cite{Sargian2003}. Momentum distributions below the Fermi momentum essentially reflect the size of the `` box  " in which the nucleons are contained. One way to model this distribution is in the simplest limit of a cluster model where a given nucleon interacts with the average potential of the other nucleons. For momenta greater than the Fermi momentum, the cluster models of nuclear structure provide enhanced strength in the momentum distribution to account for contributions with smaller spatial distributions than   the average nucleon-nucleon spacing.

Cross sections are critical observables to test ab initio calculations of nucleon momentum distributions. The large numbers of nucleon-nucleon scattering data sets, \cite{3NN}, \cite{VMC}, based on neutron-proton or proton-proton reactions, are insufficient to account for the details of nucleon momentum distributions inferred from $(e,e')$ inclusive electron scattering reactions, such as the proton-proton correlation function needed for the Coulomb sum rule, \cite{legoff}. In addition, two nucleon interactions alone cannot quantitatively explain the binding energies of low mass nuclei, \cite{3NN}, \cite{VMC}. Short range correlations between two nucleons and three body nucleon interactions are proposed to explain these  observables. $^4${He} is the best nucleus to test theoretical nucleon momentum distributions because only four nucleons are involved in this many body problem and its central density is close to that of larger nuclei.

Microscopic nuclear structure calculations based on realistic two and three body nucleon-nucleon calculations  are available for low mass nuclei~\cite{wiringa}.  In the case of $^4$He, proton momentum distributions have been calculated for proton-triton (pt) and deuteron-deuteron (dd) clusters. Recent measurements of proton-nucleon coincidences in the $^4He(e,e'pN)$ reaction~\cite{src4He,Schiavilla:2006xx,Sargsian:2005ru,Alvioli:2007zz} have shown strong correlations of back to back emission of nucleon pairs for large missing momentum $p_m$$>$400~MeV/$c$. Moreover, the increase of the  $\frac{\#pp}{\#pn}$ pair ratio as $p_m$ increases above 400~MeV/$c$ is interpreted as a sign that the nucleon-nucleon interaction is evolving from the tensor interaction to the strong repulsive short range interaction. 

\begin{figure}[thp]
 	\includegraphics[width=\linewidth]{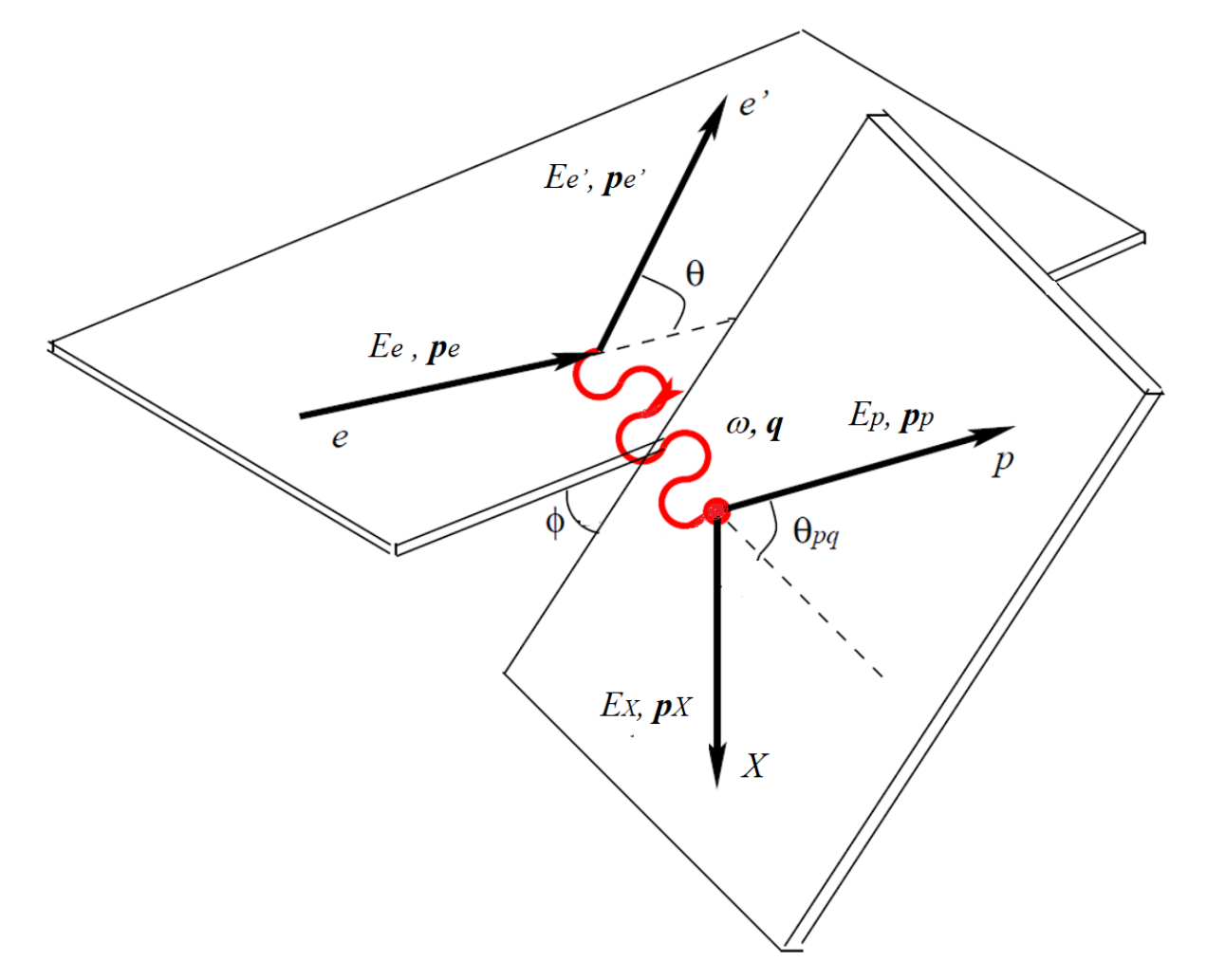}
 	\caption{Kinematic definitions for the  $A(e,e'p)X$ reaction.}  
 	\label{reaction}
 \end{figure}

Experimental access to proton momentum distributions in nuclei is possible through measurements of the differential cross section of the $A(e,e'p)X$ reaction  and its dependence on the missing momentum $p_m$ and the missing energy $E_m$. The $A(e,e'p)X$ reaction is   illustrated in Fig.\ref{reaction}, where $\vec{p}_X =\vec{p}_e-\vec{p}_{e'}-\vec{p}_p$, $p_m = |\vec{p}_X|$, is the momentum of the unmeasured particle(s) $X$ \cite{Kelly:1996hd}. The missing energy,  $E_m$, of the reaction is the difference between the electron transferred energy ($\omega$=$E_{e} - E_{e'}$) and the kinetic energies of the knocked out proton and system X, $T_p$ and $T_X$, respectively: $E_m=\omega-T_p-T_X$. The energy of the incident electron is obtained from dedicated beam energy measurements, while the energies of the scattered electron and  knocked out proton are deduced  directly from their momenta which are obtained from their respective spectrometer optics reconstruction.  The total energy of system X is obtained by the conservation of the energy in the reaction. Knowing the momentum and the total energy of X, its mass can be obtained, therefore, its kinetic energy can be obtained. 

   Previous ($e,e'p$) experiments were performed on different types of targets and as examples we cite here: \cite{Benmokhtar,Rvachev} on $^3${He}, \cite{Boeglin} on deuteron, \cite{Nilanga} and $^{16}${O}. 

Differential cross sections of proton knockout from $^4$He have a history that started with electron beam energies below 1 GeV, as in Ref. \cite{legoff}. 
But the low electron beam energy, 560 MeV,  and small duty factor(1\%), in that experiment limited the data to small $x_B$$<<$1 and missing momenta between 225 and 600 MeV/$c$. This paper provides experimental differential cross sections, in the missing energy region from 0.017GeV to 0.022GeV, called the "triton" region, based on the $^4He(e,e'p)X$ reaction over a range of  missing momenta, 25$<$$p_m$$<$632~MeV/$c$ and $x_B$$=1.24$, where X = $^3H$, $^2H + n$, and $p+n+n$, in this paper collectively called the three nucleon, 3N, mass region. 
The data were taken during the E08009 experiment in Hall A at Jefferson lab. These experimental results are compared to state-of-the-art Relativistic Distorted Wave Impulse Approximation (RDWIA) calculations of the Madrid group~\cite{udias2013} for the case X = $^3H$. However, spectra of missing energy at $p_{m} \geq 0.45$ GeV/$c$ show incursions of the X = $^2H + n$, and $p+n+n$ reactions into the "triton" region. There are no theoretical calculations available for these other final 3N states.    

 The remaining of this  paper is divided as follows: In section~II,  the E08009 experimental setup is presented, explaining the spectrometer settings and the cryogenic target. 
 Data analysis section is presented in section~III, covering background subtractions, coincidence events selection, momentum acceptance efficiency, straggling and external bremsstrahlung. 
 Details around the extraction of the cross section are presented in this section as well. In section~IV cross section results are presented, where data are compared to the Madrid group's theoretical predictions. 
 A discussion and conclusions are presented in section~V. Tables of experimental results and theoretical calculations are summarized in section VI. 
 
\section{Experimental setup}

\subsection{Spectrometer settings}

Experiment E08009~\cite{proposal} at the Thomas Jefferson National Accelerator Facility in experimental Hall A~\cite{Alcorn}, ran in February, March and April of 2011, in parallel with the triple coincidence short-range correlation experiment described in Ref. \cite{src4He}.
Data for kinematic settings of 0.153 and 0.353 GeV/$c$ missing momentum were obtained using electron beam currents between 47$\mu A$ to 60$\mu A$, for E08009. 
 In addition to these kinematic settings
 the Short Range Correlation(SRC)~\cite{src4He} experiment also obtained data at kinematic settings
out to 0.632 GeV/$c$ missing momentum including the multi-body break up channel p+X. These higher
missing momenta data were collected using 4 to 5 $\mu A$ electron beam currents but
 sufficient accumulated charge
was measured to be able to extract cross sections beyond the original goal set for E08009. Moreover,
the acceptances of the Hall A spectrometers allowed for cross sections to be determined across
a larger missing momentum range than the central value kinematic settings would suggest.

The electron spectrometer was fixed in angle and central momentum while the proton spectrometer's angles and central momenta were changed. The incident beam energy is 4.4506 GeV, the electron arm kinematic settings are as follows: electron spectrometer angle 20.3$^\circ$ electron spectrometer momentum  3.602 GeV/$c$, four momentum transfer $Q$$^2$ = 2.0 (GeV/$c$)$^2$ and Bjorken $x_b$=1.24, 3 momentum transfer of 1.647 GeV/$c$ at an angle
$\theta_q =$ 49.4$^\circ$ with respect to the incident electron momentum.  The proton arm settings are given in table~\ref{tab:central_p}.

\begin{table}
	\begin{tabular}{c|c|c|c}
		\hline
		Central $p_m$ & $\theta_p$ & $\theta_{pq}$ & Central momentum\\
		
		GeV/c  & deg. & deg. &  GeV/$c$\\
		\hline
		0.153 & 47.0 &-2.4& 1.500\\
		0.353 & 38.5 &-10.9& 1.449\\
		0.466 & 33.5 &-15.9& 1.383\\
		0.632 & ~29.0~ &~-20.4~~& 1.308 \\
		\hline
		\end{tabular}
		\caption{Proton spectrometer settings}
		\label{tab:central_p}
		\end{table}

\subsection{Cryogenic target}
The cryogenic target was gas $^4$He contained in an aluminum can of length 20~cm. The nominal temperature
of the gas was $20^\circ$~K at 199~psia. $^4$He enters and exits at the upstream end of the target. There is no outlet for the fluid at the downstream end of the can. A determination of target density along the beam path was done by comparing the normalized yield of scattered electrons at  
$47\mu A$ and $60\mu A$ beam currents to the yield at $4\mu A$. Since the electron spectrometer was
held at a fixed momentum and angle the electron spectrometer served as a density monitor. For this
target at a beam current of $4\mu A$ a computational fluid dynamics (CFD) calculation~\cite{cfd} predicts
an average density drop of 2.3\% from strictly thermodynamic parameters. A comparison of the measured
yield at $4\mu A$ to the CFD calculation gives an uncertainty in the target density dependence along the
beam of 1.1\%. More detail for the treatment of the target density used in the data analysis is available in 
~\cite{sana}. Across the  $\pm$~8cm effective target length and for the different beam currents, the target densities are summarized in table \ref{tb1:DensityCalculation}.

\begin{table}[htbp]
	\begin{center}
		\begin{tabular}{c|c}
			\hline
			Beam current & Target density  \\
			($\mu$A)	 & $nuclei/cm^2$\\
			\hline
			4.014  &  $7.84 \pm 0.087 \times 10^{22}$  \\
	    	45.46 &   $6.732 \pm 0.077 \times 10^{22}$ \\
		    60.71 &  $5.662 \pm 0.065 \times 10^{22}$ \\
		\hline
	\end{tabular}
\end{center}
	\caption{Target density dependence on beam heating as a function of beam current.} 
\label{tb1:DensityCalculation}
\end{table}


\section{Data Analysis}

\subsection{Background subtraction and coincidence event selection}
\begin{figure}[htp]
	\includegraphics[scale=0.35]{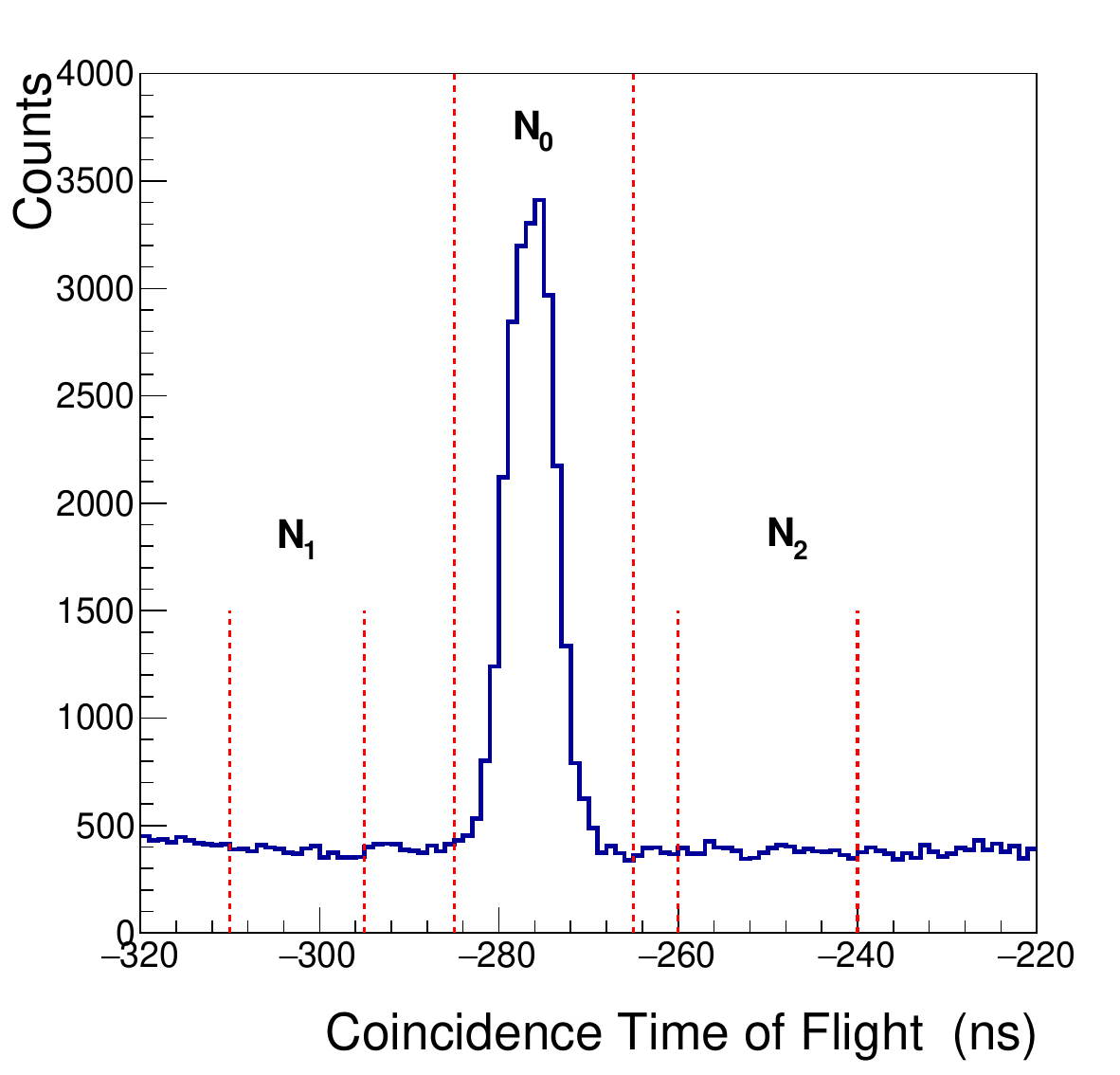}
	\caption{Coincidence time of flight spectrum for the 353 MeV/$c$~ setting.}  
	\label{CoincTime}
\end{figure}

For this experiment, event triggers were generated by coincident signals from
scintillator arrays. Particle tracks were reconstructed using
the high resolution spectrometer's  vertical drift chambers. The small $\pi^-$ background
in the electron arm was rejected using a CO$_2$ gas Cherenkov detector. In the proton spectrometer, coincident $\pi^+$, and other positively charged nuclei like $^2$H, and $^3$H  were separated from the protons using the time difference between particles detected in the two spectrometers. 
Most of the accidental coincident events were rejected by cuts on the difference between interaction points in the target along the beam as reconstructed by the two spectrometers. The remaining accidental background was subtracted using the coincidence timing between the spectrometers. Fig.~\ref{CoincTime} shows a coincidence time of flight for the 353 MeV/$c$~kinematics. The number of real coincidence events in a 20~ns time window around the peak was obtained by subtracting the accidentals under the peak considering a flat background under the whole spectrum, as shown in Fig. \ref{CoincTime}.

\begin{figure}
	\includegraphics[scale=0.35]{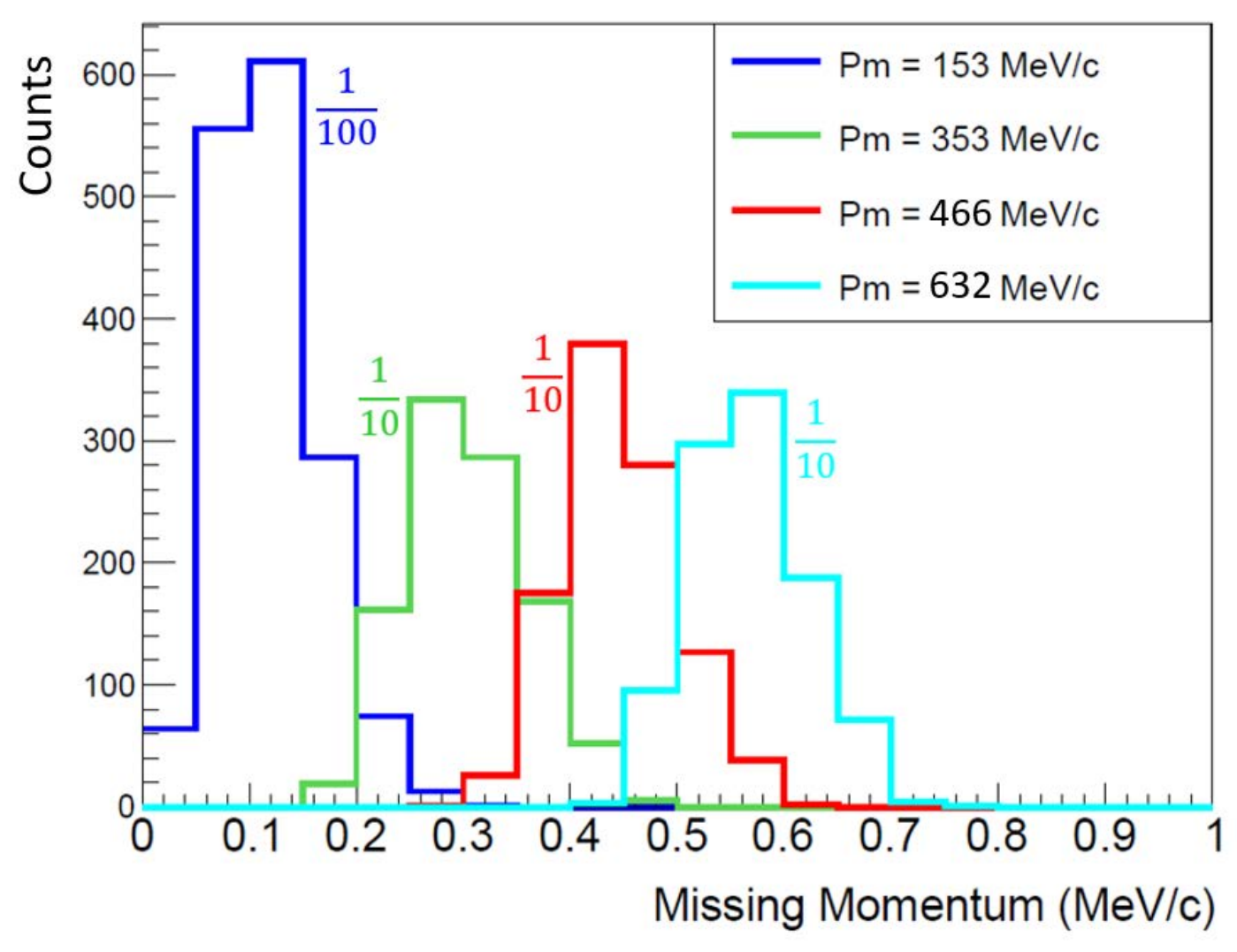}
	\caption{Missing momentum spectra for all the kinematical settings of the experiment.}  
	\label{Momenta}
\end{figure}

The wide momentum acceptance of the spectrometers allows for a broad missing momentum acceptance as shown in Fig. \ref{Momenta}, so we were able to divide the study in 50 MeV/$c$ wide bins in $p_m$.
 For each kinematical bin, the number of true coincidence events $N_t$ was determined from the coincidence time of flight with the formula:

\begin{equation}
N_t=N_{0}-\frac{\Delta {t_0}(N_1+N_2)}{\Delta {t_1}+\Delta {t_2}}
\end{equation}

where $N_0$ is the number of events within the bin reconstructing in
the real coincidence window $\Delta{t_{0}}$, and $N_1$ and $N_2$ are
the number of events within the bin reconstructing in the accidental
coincidence windows $\Delta{t_{1}}$ and $\Delta_{t_{2}}$,
respectively. Statistical uncertainties were propagated as

\begin{equation}
\delta {N_t}=\sqrt{N_r+(N_1+N_2)}{\left( \frac{\Delta {t_0}}{\Delta{t_1}}+\Delta {t_2} \right)}^2
\end{equation}

\begin{figure}
 		\includegraphics[scale=0.21]{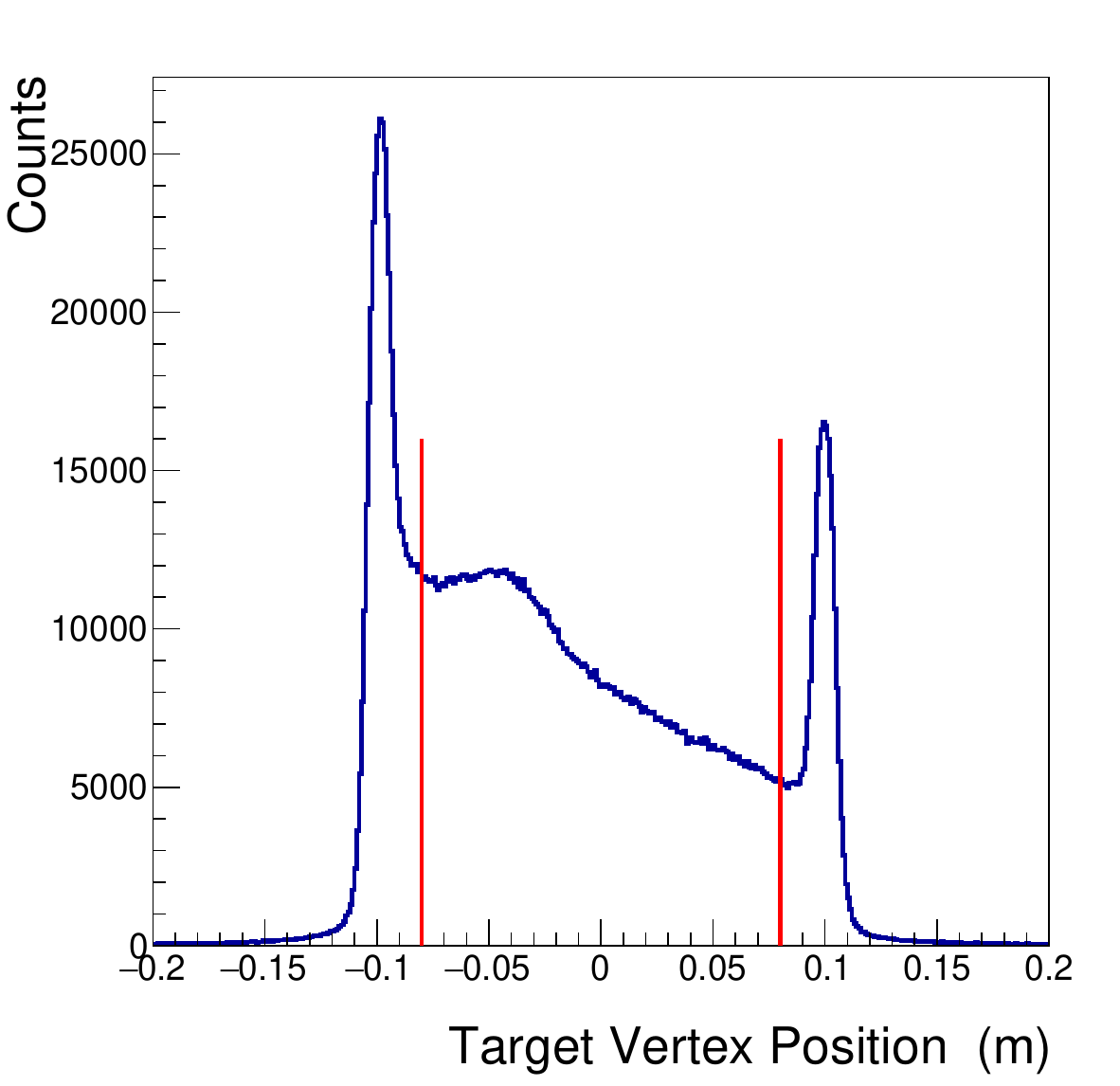}  
 	\includegraphics[scale=0.21]{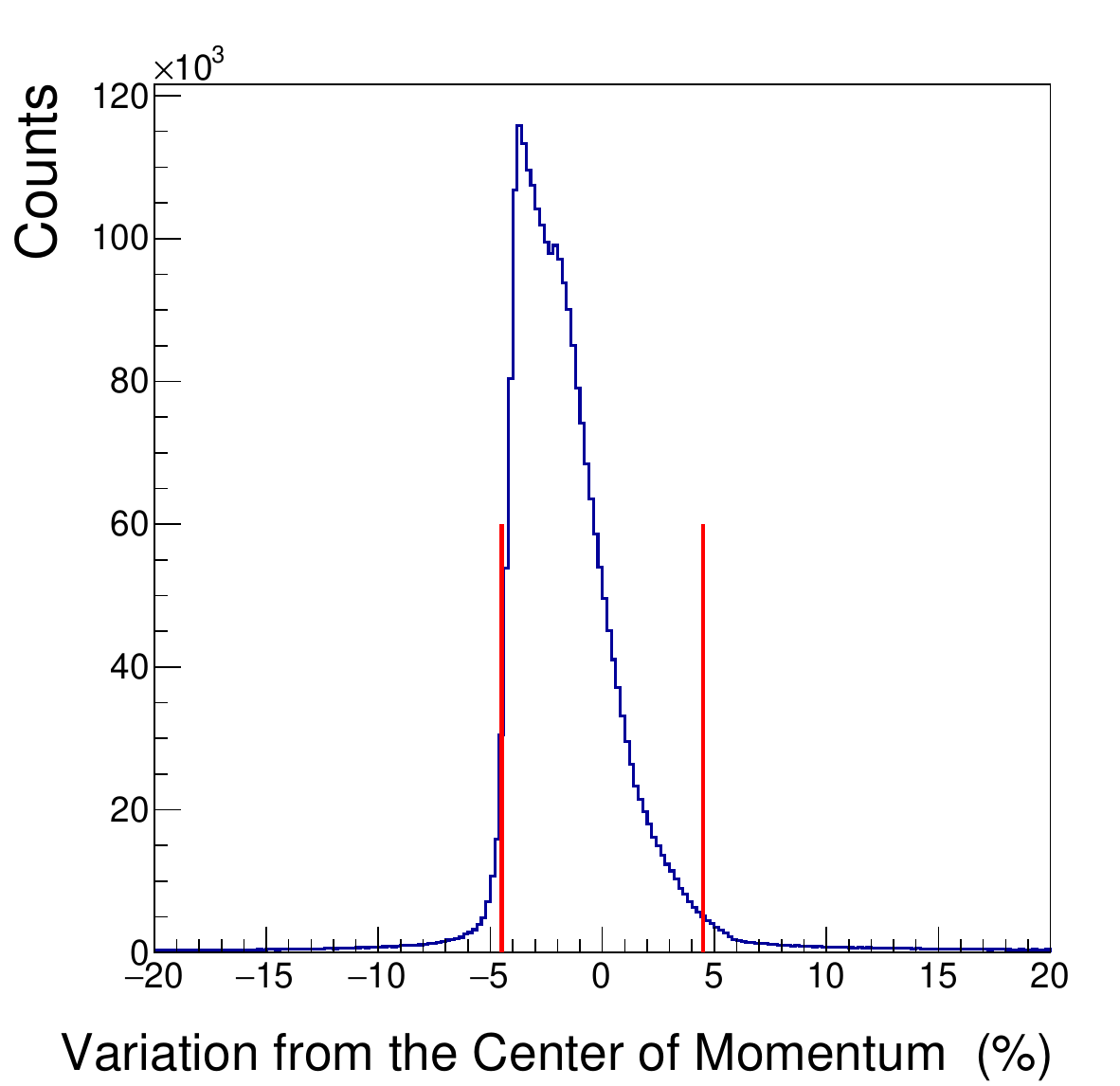}
 	\includegraphics[scale=0.21]{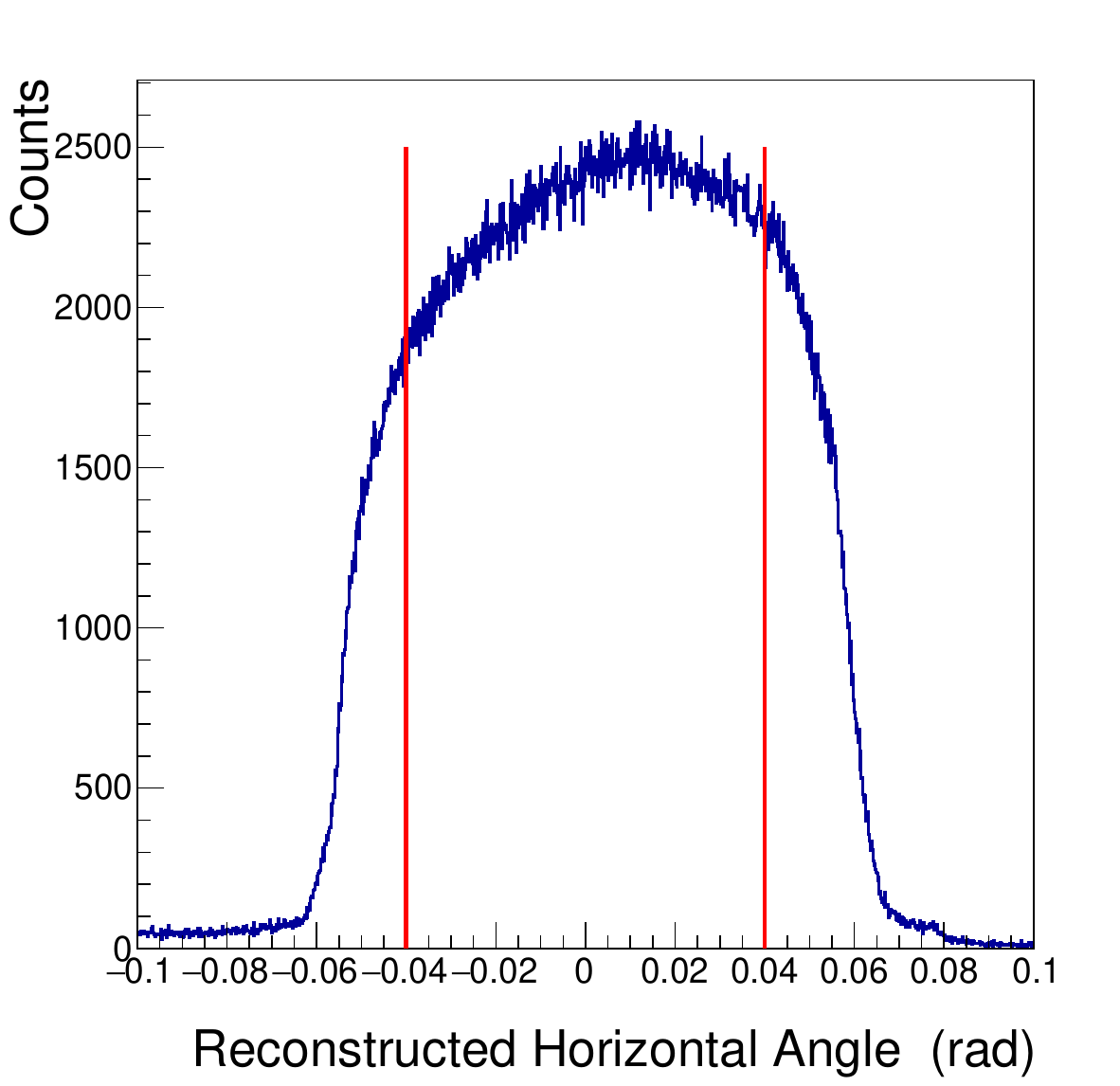}
 	\includegraphics[scale=0.21]{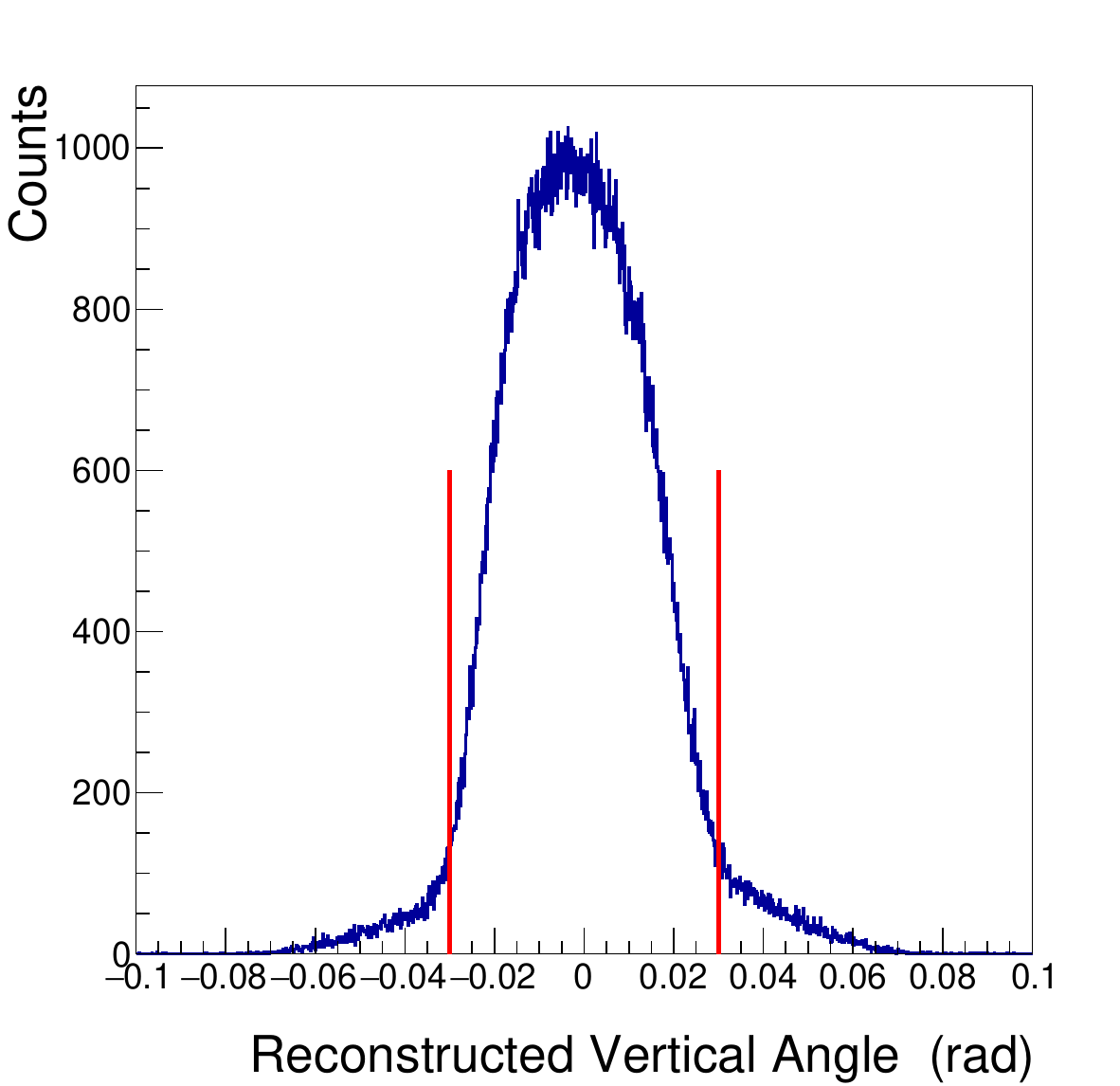}  

	\caption{upper left: Target vertex position in meters, upper right: Variation from the central momentum \%.  bottom left: reconstructed horizontal angle in radians, bottom right:Reconstructed vertical angle in radians.} 
	\label{Acceptance}
\end{figure}

For the determination of the cross section, the following phase-space cuts are applied to the data for both electron and proton spectrometers: horizontal angle $ \pm 0.04$ radians,
vertical angle  $\pm 0.03$ radians, vertex position $\pm 8$~cm and the deviation from central momentum $\pm 4.5\%$. These variables are shown in Fig.\ref{Acceptance}.

\begin{figure}
	\includegraphics[angle=0,scale=0.5]{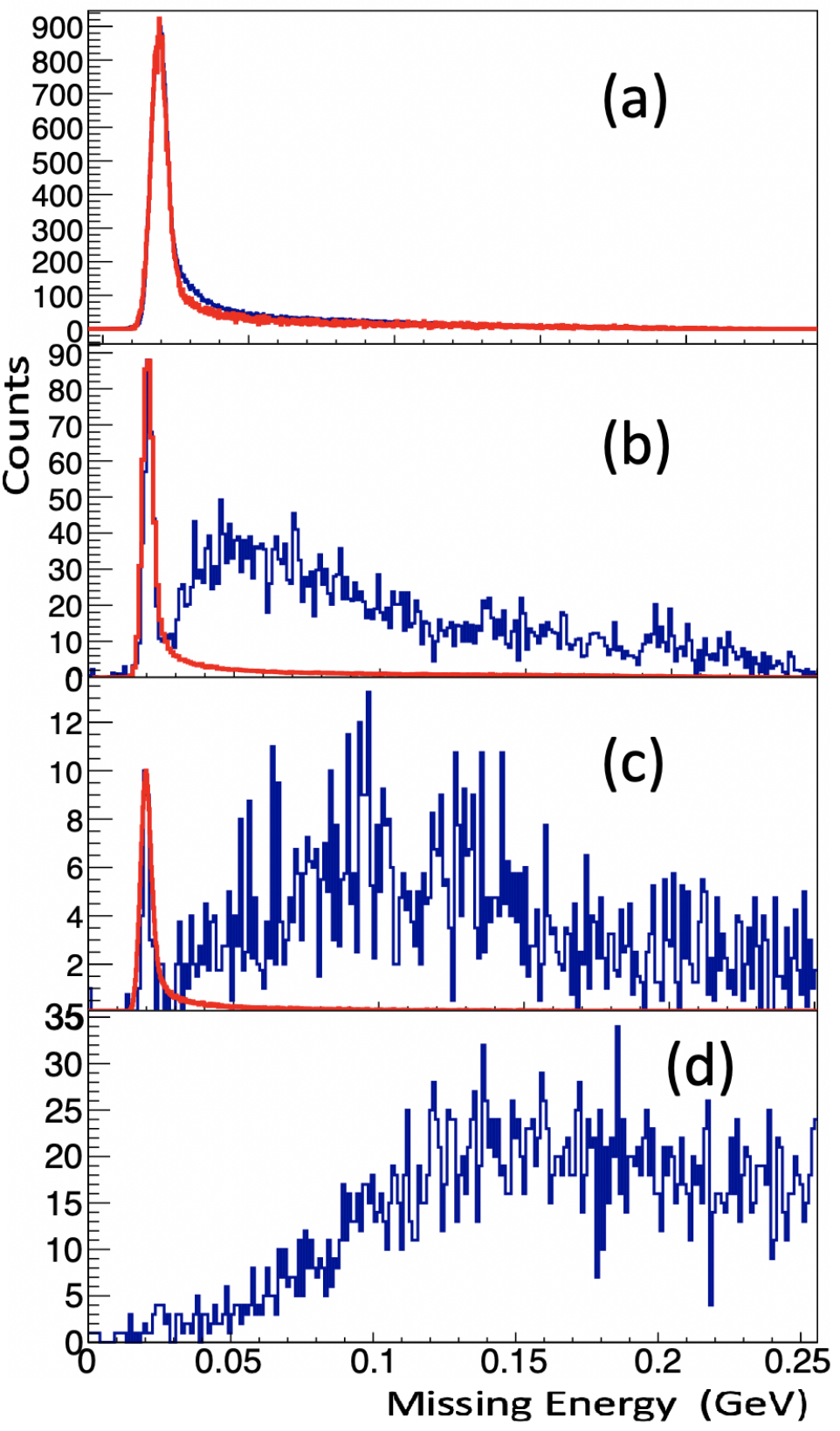}
	\caption{Missing Energy spectra for all the kinematical settings. Data are in blue and the simulated two body breakup channel is in red. From top to Bottom: (a) $p_m$ = 153 MeV/$c$, (b) $p_m=$353 MeV/$c$,  (c) $p_m$=466 MeV/$c$ and (d) $p_m$=632 MeV/$c$}  
	\label{fullData}
\end{figure}

\begin{figure}
	\includegraphics[scale=0.45]{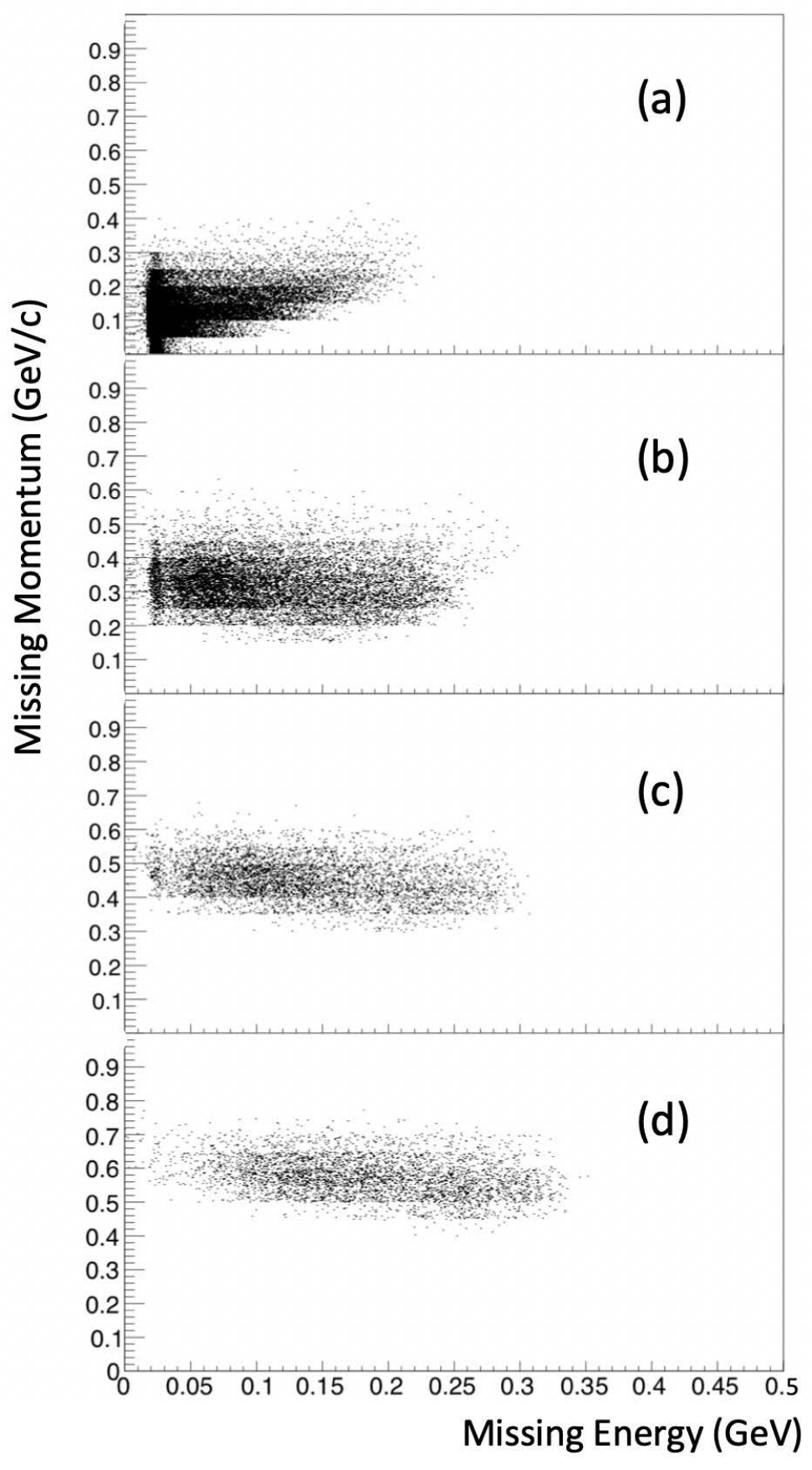}
	\caption{Two dimensional plots of missing momentum versus missing Energy for the kinematical settings of the experiment. From top to Bottom: (a) $p_m$ = 153 MeV/$c$, (b) $p_m=$353 MeV/$c$,  (c) $p_m$=466 MeV/$c$ and (d) $p_m$=632 MeV/$c$}  
	\label{EmissPmiss}
\end{figure}

Missing energy spectra for all the kinematics after accidental and background subtraction are presented in figure \ref{fullData}. Figure \ref{EmissPmiss} represents two dimensional missing momentum versus missing energy spectra for the full data set. Note that strength  of the two body cluster weakens while going from lower to higher momenta.    

 Data analysis is aided by the Monte Carlo simulation (GEANT 3.2~\cite{geant32}) of the transport
 of the incident electron, scattered electron and proton through the target cell  into the spectrometer apertures, assuming a p+triton final hadronic state. The identification of the p+triton final state is possible by calculating the missing energy in the scattered electron + p state. A peak in the missing energy spectrum, at 19.8 MeV, corresponding to the triton ground state identifies the $^4He(e,e'p)^3H$ reaction, as seen in figure~\ref{fullData}.

 \subsection{Missing momentum acceptance efficiency}\label{momentum}
 
  In the simulation a vertex point is chosen at random in the long gas target which gives the incoming electron's momentum at the interaction point. Then hit points within the apertures of the spectrometers for the outgoing electron and proton are randomly selected. Each point within the spectrometers' apertures has an equal probability of being selected. This allows for the vertex angles of the electron and proton
 to be determined. An energy for the outgoing electron is chosen within the momentum  acceptance of the electron spectrometer. From the incident electron's momentum, the scattered electron's momentum and the angles for the ejected proton,  three body kinematics for the $^4He(e,e'p)^3H$ reaction allows for the proton's vertex momentum to be
 determined. The electron and proton are followed from the vertex to the final hit points in the spectrometers' apertures. Thus complete information about the location and momenta at the vertex and the spectrometers' apertures is known.
 
 The three body kinematical and geometrical limitations for particles arriving at the hit points within the apertures are calculated by GEANT and thus allows the missing momentum, $\vec{p_m} = \vec{p_e} - \vec{p_{e'}} -\vec{p_p}$ to be calculated.  In the analysis we bin $|\vec{p_m}|$ into 50~ MeV/$c$ bins and we define the missing
 momentum acceptance factor, $f(p_m)$, for a bin as:
 \begin{equation}
 f(p_m)=\frac{n(p_m)}{\sum n(p_m)}.
 \end{equation}
 
 where $n(p_m)$ is the number of triton events in the missing momentum bin centered
 on $p_m$ and $\sum n(p_m)$ is the total number of triton events over all missing
 momenta for the particular proton kinematic setting. The same Gaussian broadening
 used for the simulation fit in figure~\ref{fullData}(b) is used to generate the
 values of $p_m$ needed to calculate $f(p_m)$.\\
 
 The momentum resolution measured by the spectrometers in this experiment is a factor
 of 3 to 4 times larger
 than for a point target due to the 16 cm of target length we used. This shows up in
 the width of
 the missing energy spectra. We see from figure 5(b) a strong peak near the triton
 ground state and the
 background from other processes. When we compare our data to theory we are
 restricted to a window around
 the peak. However, the theoretical calculations usually do not include the
 scattering and radiative
 effects seen in the data.
 
 The missing momentum factor is our estimate of how many of the theoretical events
 fall outside our experimental window. There is a systematic uncertainty in this
 factor because it is only 
 calculated by GEANT. Ideally, we want to use an independent experimentally determined
 missing momentum factor 
 established on a well known data set of ($\it{e,p}$) coincidences. However, such an 
 experimental data set is not
 available over the full range of electron and proton momenta measured in this
 experiment. We were encouraged
 to see that this simple choice of missing momentum factor follows the theoretical
 predictions quite well, see figures~\ref{fig:data_full} and~ \ref{fig:ratio}.
 While this does not give a precise calibration of the acceptance, we estimate that the systematic uncertainty in $f(p_m)$ is 10$\%$, which is conservative given the typical uncertainty in the acceptance when physics checks can be performed.  
 %
 \subsection{Peak broadening effects}\label{radiative}
 Straggling and external Bremsstrahlung obtained from the GEANT simulation  produce a broadening and a
 characteristic tail on the missing energy spectrum. In practice the long
 target introduces additional broadening beyond the intrinsic point source resolution
 of the spectrometers. 
 The additional broadening is included in the simulation
 by a Gaussian smearing of the momenta at the apertures. It is typically a factor of three to four bigger than the resolution of the point source peak. The amount of Gaussian smearing needed is determined by the best fit of a strong missing energy data peak such as at the lowest missing momentum. An example of the fit is seen in figure~\ref{fullData}(b) where the simulation of the two body break-up channel is represented in red. \\

  \subsection{Extraction of the Cross Section }
 
 The average cross section for  the $^4He(e,e'p)X$ reaction per missing momentum bin was extracted for the triton region and it is given by: 

\begin{equation}
<\sigma(p_{m})>=\frac{n(p_{m})*RSC}{\Delta\Omega_{e}\Delta\Omega_{p}\Delta E_{e}N_{e}N_{tgt}*Eff}.
\end{equation}


where:
\begin{itemize}
	\item[a)] $n(p_{m})$ is the net counts in the triton region between missing energies of 0.017 GeV to 0.022 GeV, after randoms and background subtraction. Since there is no model for the X = 3N channels beyond 0.022 GeV, and since these channels reach 0.029 GeV, the background subtraction in the triton region was done using straight line subtraction below 0.029 GeV. An example of this background subtraction for 153~MeV/$c$ kinematics is shown in Fig. \ref{Emissdirty}. Left and right plots are before and after background subtraction, respectively. The net count in the triton region is obtained by the total counts in the shaded area in the right plot. 
	
	\begin{figure}[htp]	
		\includegraphics[scale=0.34]{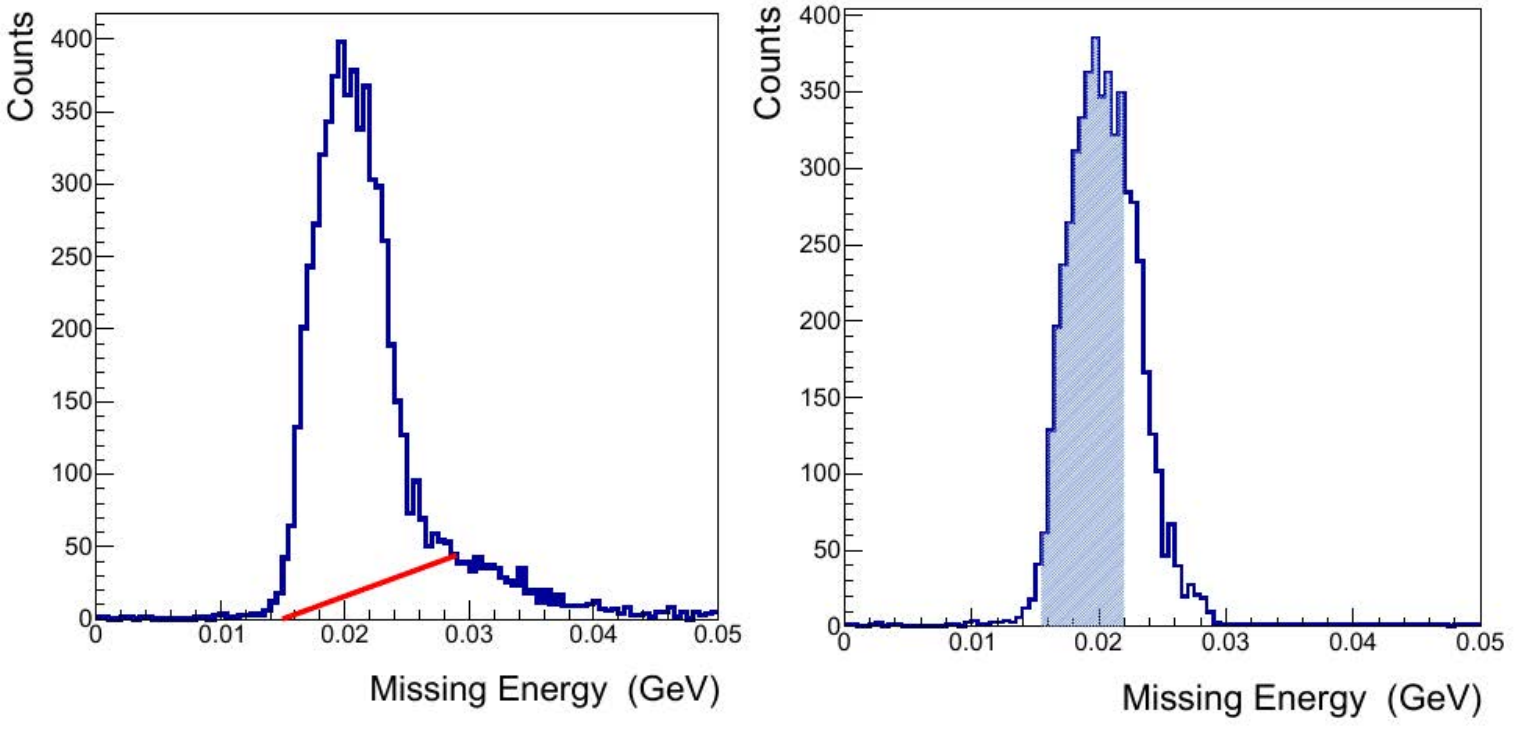}
		\caption{Missing energy spectrum for the 153 MeV/$c$ kinematics. Left: before straight line background subtraction and showing the position of the line in red.  Right: After the straight line background subtraction.}  
		\label{Emissdirty}
		
	\end{figure}
		
	\item[b)] $RSC$ is the radiative and straggling corrections to the cross section due to the tail on the missing energy spectrum.  These corrections are determined by comparing the number of events in a 5~MeV window centered on the triton peak to the total number of events in the GEANT simulation. There is little variation in RSC from the simulation between proton spectrometer settings: 1.33$<$RSC$<$1.35. There is an uncertainty of 0.2\% on RSC.
	
	
	\item[c)] $\Delta\Omega_{e}$ and $\Delta\Omega_{p}$ are the geometrical solid angles of the spectrometer apertures. 
	
	\item[d)]  $\Delta$$E_{e}$ is the electron momentum bin in coincidence with protons. The choice of the bin size is determined by the proton arm by studying the dependency of the proton momentum versus scattered electron momentum. A two dimensional plot of  proton momentum versus electron momentum for the total coincidence events is presented in  Fig.\ref{PpPe} for the 153 MeV/$c$ kinematics. Plots for higher momenta look similar with less statistics. 
	  This plot was then studied for various missing momentum bins for a given kinematics. This reduces the statistics making the choice of $\Delta$$P$ somehow ambiguous. Therefore, a systematic uncertainty of $10\%$ was attributed to $\Delta$$E_{e}$.
	
	\begin{figure}[htp]
		\includegraphics[scale=0.35]{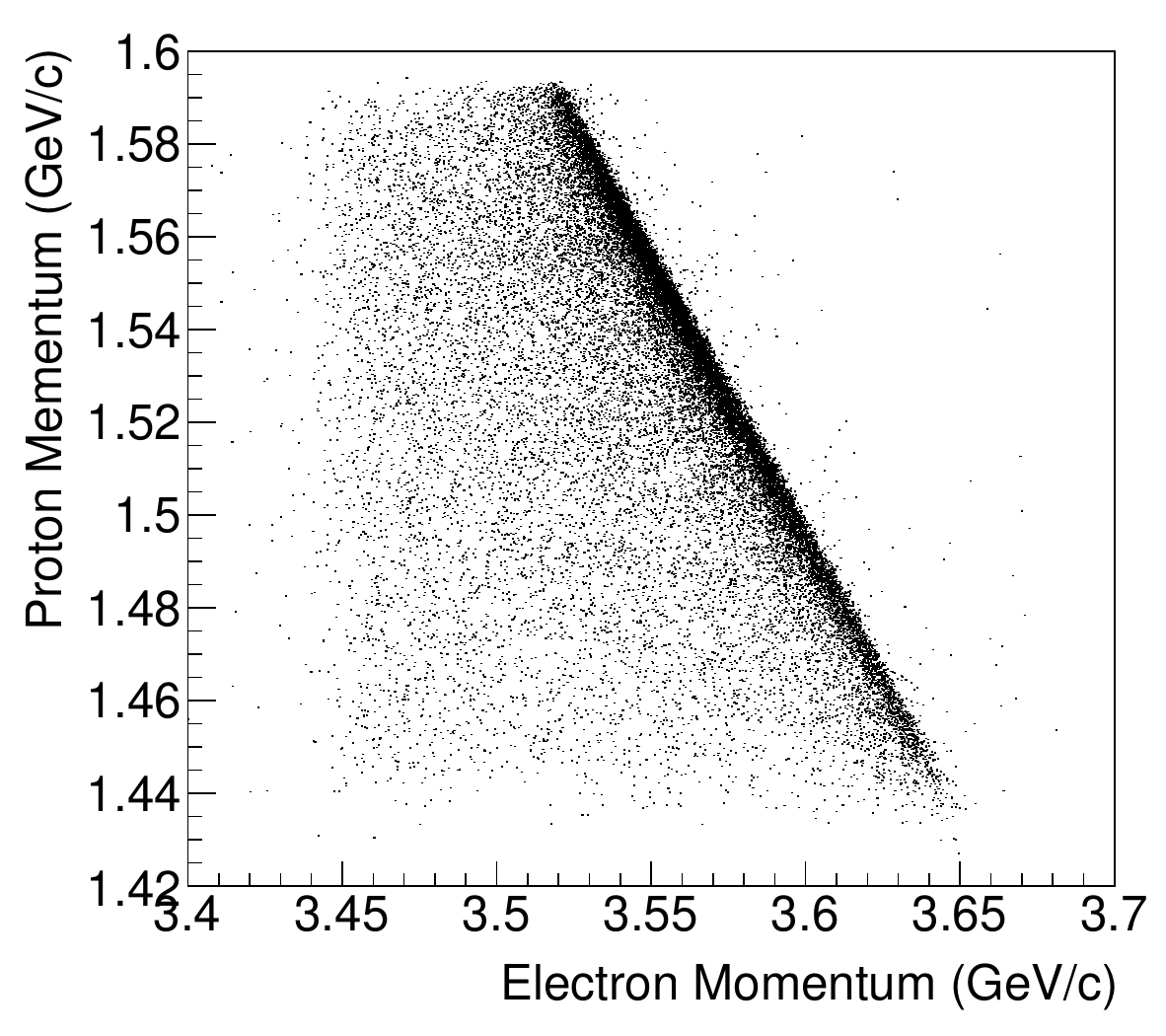}
		\caption{Proton momentum versus electron momentum for the 153 MeV/$c$ kinematics.}  
		\label{PpPe}
	\end{figure}

	
	\item[e)] $N_{e}=Q/e$ is the number of electrons that passed through the target, where $e$ is the electron charge and $Q$ is the total charge. This is measured by the beam current monitors with an uncertainty of 0.3-0.5$\%$. 
	
	\item[f)] $N_{tgt}=\rho(I) * z_{tgt}$ is the number of nuclei per cm$^{2}$ in the beam. $I$ is the beam current, $\rho(I)$ is the number of nuclei per cm$^{3}$ and $z_{tgt}$ is the effective target length. Target densities along the {$\pm$8~cm} effective target length for different beam currents are presented in table \ref{tb1:DensityCalculation}. $N_{tgt}$  was known to 1.14\%.   
	\item[g)] $Eff$ is the efficiency factor and it accounts for:
	\begin{enumerate}
	 \item the missing momentum acceptance factor: $f(p_m)$ that is explained in \ref{momentum}. 
	 \item data acquisition live time (LTdaq), 
	 \item electronics live time (LTel),
	 \item trigger efficiency (Tri), 
	 \item wire chamber (WC) and
	 \item tracking efficiencies (Tra).
	 \end{enumerate}
 
	  This efficiency  is given by: 
	 \begin{equation} 
	 Eff=f(p_{m}) \cdot LTd \cdot LTe \cdot Tri \cdot WC \cdot Tra 
	\end{equation} 
	
	 \end{itemize}

	 	
	 	
	  The live time of the trigger acquisition system, $LTdaq$, was $0.916 \pm 0.01$,
   and $0.95 \pm 0.01$ for the 153~MeV/$c$ and 353~MeV/$c$ kinematics, respectively. For the higher missing momentum settings, $LTdaq$ was larger than 0.99, with negligible uncertainties. The remaining efficiencies are displayed in table \ref{tb1:GeneralUncertainties}.
   

\begin{table}[htbp]
\begin{center}
	\begin{tabular}{c|c|c}
		\hline 
		Efficiency & value & Uncertainty (\%)\\
		\hline
		Electronic live time & 1  & 0 \\
		Trigger efficiency & 0.97 & 1 \\ 
		Wire chamber efficiency & 0.995 & 0.1 \\ 
		Tracking efficiency & 0.9895 & 0.75 \\
		\hline
	\end{tabular}
\end{center}
\caption{General uncertainties}
\label{tb1:GeneralUncertainties}
\end{table}

To conclude this analysis section, a summary of the parameters involved in the cross section calculations and their uncertainties are presented in section~\ref{cross} table~\ref{tab:uncertainties}.
   Experimental differential cross sections, $\frac{d
			\sigma^5}{d\Omega_p d\Omega_e dE_e}$,
		for $^4He(e,e'p)X$; where  $X$ $=$ $^3H$ or $3N$,  from E08009, for
		different kinematical settings given by the proton spectrometer central angle are presented in table~\ref{tab:xsect}. Statistical
		and normalization uncertainties are the first
		uncertainty entry, systematic uncertainties in selecting the size of the $\Delta E$ bin of 10\% and an estimated 10\% from the missing momentum acceptance factor, described in sections IIIB and IIID(d) are the second uncertainty entry, total uncertainty is the third entry. The total uncertainly is calculated by adding statistical and normalization uncertainly and the systematic uncertainty in quadrature.



\section{Results}

\subsection{Comparison of data to theoretical predictions}

The extracted differential cross sections are compared to relativistic distorted wave impulse approximation
calculations of the Madrid theory group ~\cite{{udias2013},{malace},{3heeep},{udias1}}.
The $^4$He ground state is described by a relativistic solution of the Dirac equation phenomenologically adjusted to
fit the observed radius and binding energy of $^4$He. These calculations were first introduced in ~\cite{Paolone}.

Vertex values of the incident electron's momentum at various positions
within the long gas target and the momenta of the scattered electron and
ejected proton were provided to the Madrid theory group for calculation of the cross section
at each event vertex in the GEANT simulation. The GEANT simulation also contains the
detected electron and proton momenta at the spectrometers' apertures. In this way the
vertex cross section can be associated with the missing momentum at the apertures.

\begin{figure}[htp]
\includegraphics[scale=0.4]{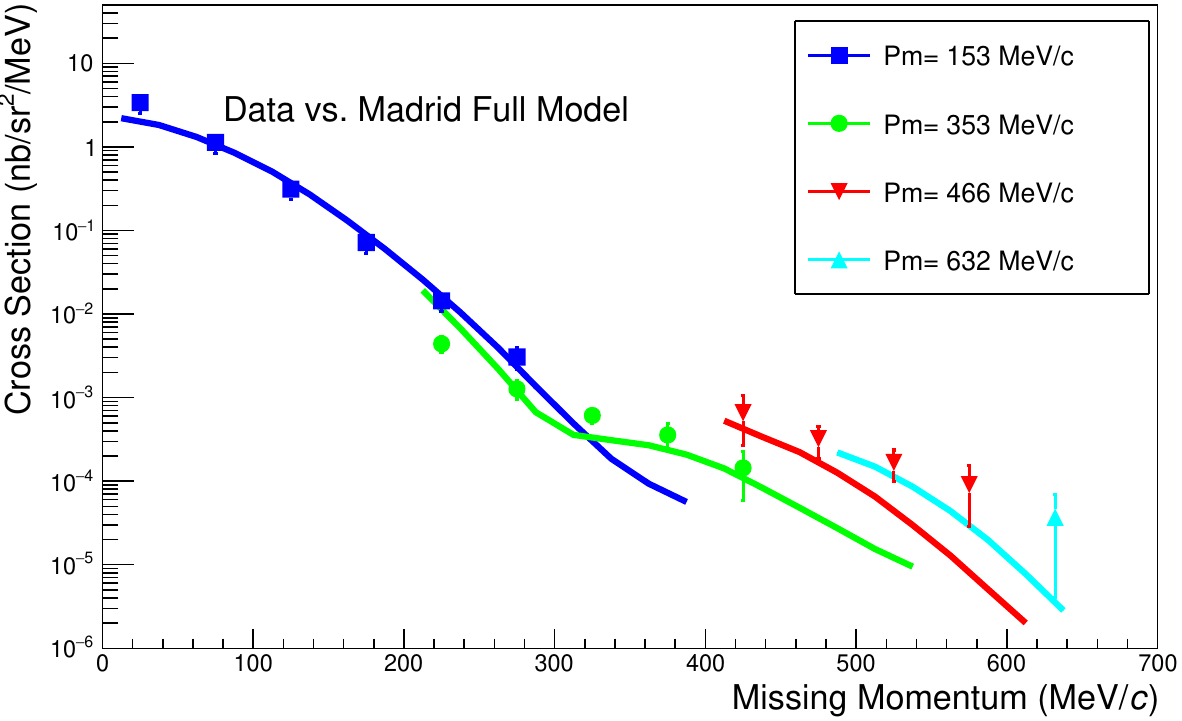}
\caption{E08009 Data compared to Madrid full theoretical calculations. Blue squares are for the 153 MeV/$c$ setting, green circles are for the 353 MeV/$c$ setting, red inverted triangles are for the 466 MeV/$c$ setting and cyan triangles are for the 632 MeV/$c$ setting. Theoretical calculation follow the same color code as the data for each momentum setting.}
\label{fig:data_full}
\end{figure}

\begin{figure}[htp]
\includegraphics[scale=0.4]{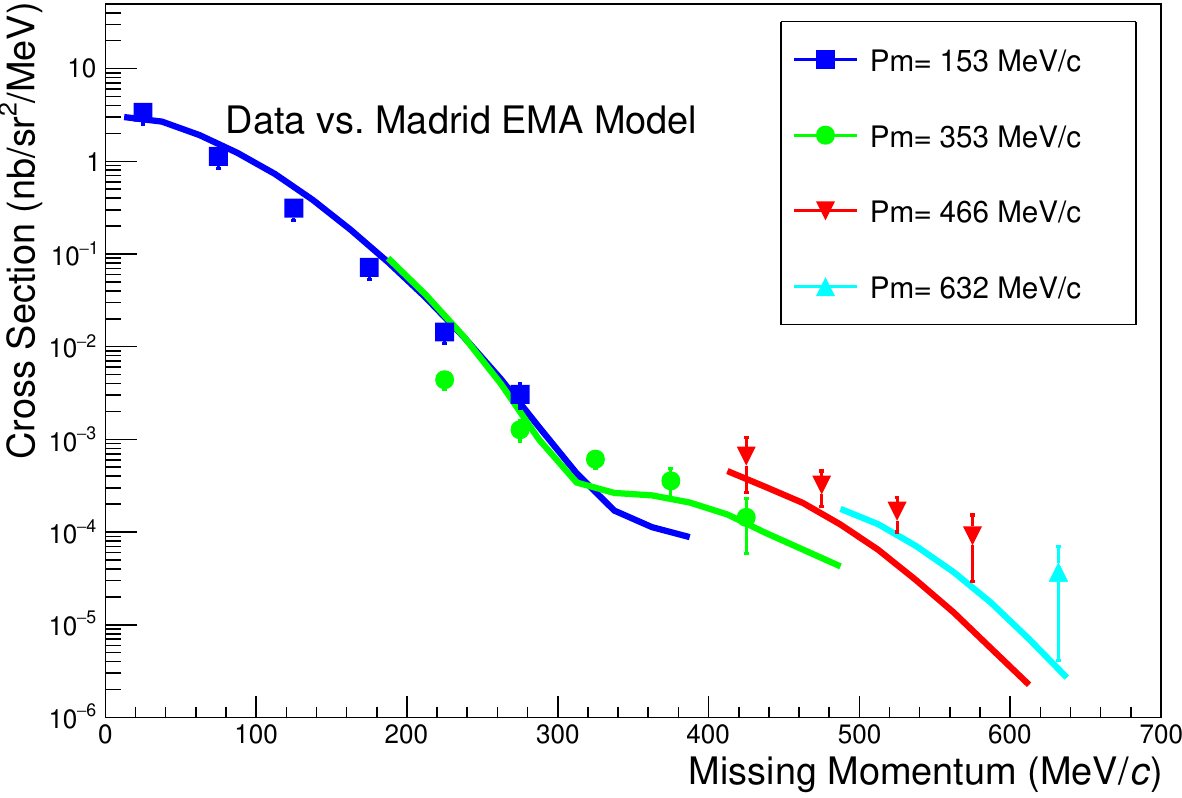}
\caption{E08009 Data compared to Madrid EMA theoretical calculations.  Blue squares are for the 153 MeV/$c$ setting, green circles are for the 353 MeV/$c$ setting, red inverted triangles are for the 466 MeV/$c$ setting and cyan triangles are for the 632 MeV/$c$ setting. Theoretical calculation follow the same color code as the data for each momentum setting.}
\label{fig:data_EMA}
\end{figure}

Theoretical cross sections integrated over the experimental acceptances for
the full Madrid treatment and using the effective momentum approximation, EMA, treatment are presented in tables~\ref{tab:madridfull} and
~\ref{tab:madridEMA}. Plots of the data for the two theoretical treatments are shown in
figures~\ref{fig:data_full} and ~\ref{fig:data_EMA}. For these kinematics, the EMA calculation is nearly indistinguishable from the full calculation, except for small differences at very low missing momentum.


Data and calculations show the same missing momenta dependence for the measured or calculated cross 
section as a function of kinematic setting. Even though the same magnitude of $p_m$ is reached
for different proton angles the cross section does not simply factor as a function of $p_m$.  Good agreements between the Madrid calculation and the data extend to about 420~MeV/$c$ in missing momentum. It can be also noticed that both data and theory exhibit an inflection in the slope of the cross section between 300 and 400~MeV/$c$. In recent calculations on light nuclei~\cite{wiringa}, an inflection in the proton momentum distributions was predicted in the momentum range between 0.2 and 0.6~GeV/$c$. For  $^4$He, this inflection appears to be due to the triton+proton cluster distribution exhibiting a deep minimum in the proton momentum distribution. When added to the deuteron-deuteron cluster distribution, the inflection appears below and close to 0.4~GeV/$c$ in the total proton density distribution, which is in agreement with the one we see in these data. 


\section{Discussion and Conclusion}

For this experiment, the momenta of the outgoing proton and scattered electron in the  $^4He(e,e'p)X$ reaction are measured. Using energy and momentum conservation, we can determine the momentum of the undetected hadronic state $X$. Theoretical comparison to the data here is limited to a specific exit channel, $X$ = $^3$H. However, considering the 
theoretical cluster contributions to the proton momenta~\cite{wiringa} in $^4$He, the contribution of the $pt$ cluster to the proton momentum distribution is expected to be negligible above about $p_m$ =250~MeV/$c$.

\begin{figure}[htp]
	\includegraphics[scale=0.48]{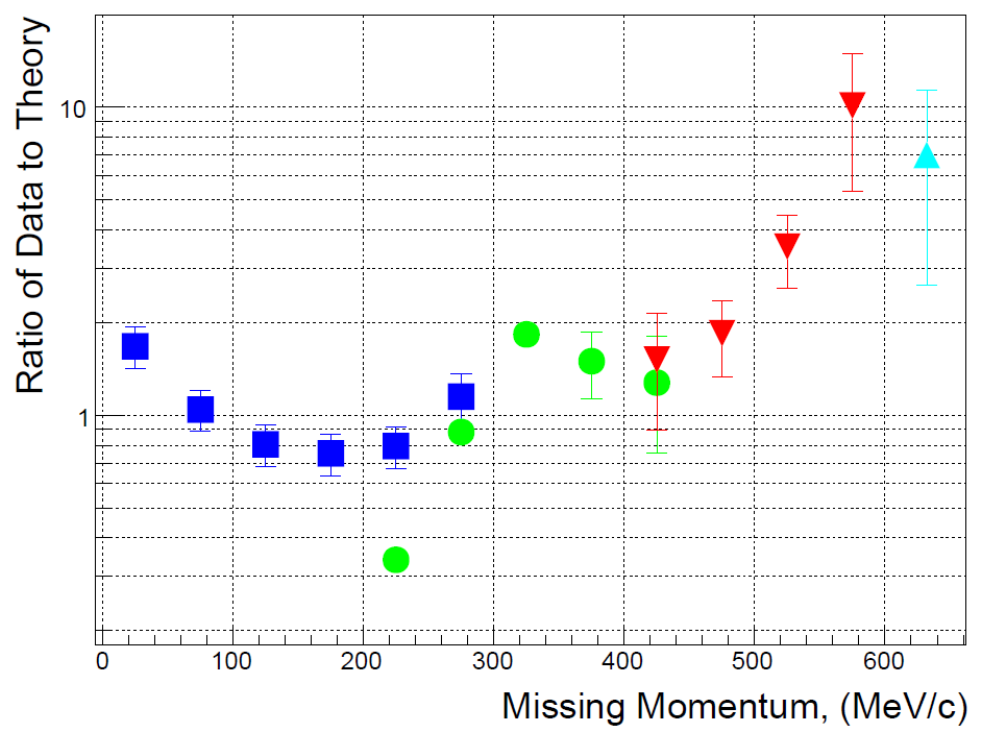}
	\caption{Ratio of the experimental cross section to the theoretical Madrid full (pt) cross section versus missing momentum. Squares are for the 153 MeV/$c$ setting, circles are for 353 MeV/$c$ setting, inverted triangles are for the 466 MeV/$c$ setting and triangles are for the 632 MeV/$c$ setting.}
	\label{fig:ratio}
\end{figure}

The ratio of experimental cross section to the Madrid full predictions, in logarithmic scale, is
shown in figure \ref{fig:ratio} for the four proton spectrometer central momentum settings. The blue squares, at the lowest missing momentum setting, hover around a ratio of 1, showing good agreement between data and predictions. The green dots are for the 0.353 GeV/$c$ setting and we see a distinctive pattern for these data. The ratio at 0.225 GeV/$c$ is 0.34, substantially different from the model prediction. This behavior cannot be traced to a statistical fluctuation because as we see in figure \ref{fullData}(b), there is a substantial
peak at the triton missing energy location. The cross section decreases by a factor of 12 between 0.225 and 0.325 GeV/$c$ and over the full range in missing momentum for this proton angle setting the cross section falls by a factor of 30. This fluctuation of the data to theory ratio suggests that some significant physics is not adequately included in the theoretical model for this range of missing momentum with these spectrometer settings. For the data at the 0.466 and 0.632 GeV/$c$ settings the ratio again shows a smooth missing momentum dependence.

However, the overall dependence of the cross section by the Madrid full model in figure \ref{fig:data_full} is qualitatively described. 

From~\cite{wiringa}, the high proton momentum is attributed to the repulsive nucleon-nucleon core.
Fig.\ref{fullData} shows a broad peak in the missing energy spectrum which shifts in position kinematically with the photon being absorbed on a correlated pair of nucleons. This feature has been previously seen in $^3He(e,e'p)pn$ measurements in Ref. \cite{Benmokhtar} and \cite{marchand} and in $^4He(e.e'p)X$ continuum channel in Ref.\cite{legoff}.

The measurements of ~\cite{src4He} are consistent with the NN short range force becoming repulsive. However,
it is counter intuitive and in disagreement with theoretical expectations ~\cite{wiringa} that tritons
should be ejected from $^4$He along with protons emerging from short range encounters.

 The fact that we observe events in the triton region up to $p_m$ = 632~MeV/$c$ involves processes beyond the impulse approximation.
 Final state interactions of the outgoing proton may take a proton knocked out of a pt cluster initially at a low value of $p_p$ to appear as if its momentum at the vertex was $p_m$. This is accounted for to some extent by the optical model potential treatment of the final $pt$ unbound state. We see good agreement between the theory and data in figure~\ref{fig:data_full} up to about $p_m$=420~MeV/$c$.  
 
Beyond about 450~MeV/$c$ in $p_m$ substantially more triton region events are measured than what the Madrid full theory predicts. In this case three nucleons emitted at high $p_m$ may be a signature of other reactions allowing the three nucleon
cluster to emerge as a bound or quasi bound state. Since the kinematics for the electron were chosen for $x_b=$1.24, protons
in more intimate interactions with neighbors than quasi-elastic conditions($x_b \approx 1$) may favor
other reactions leading to three nucleon clusters exiting in the missing energy region near the triton.

\begin{figure}[htp]
	\includegraphics[scale=1.14]{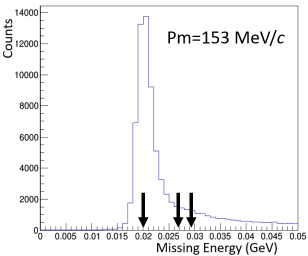}
	\label{fig153}
		\includegraphics[scale=0.4]{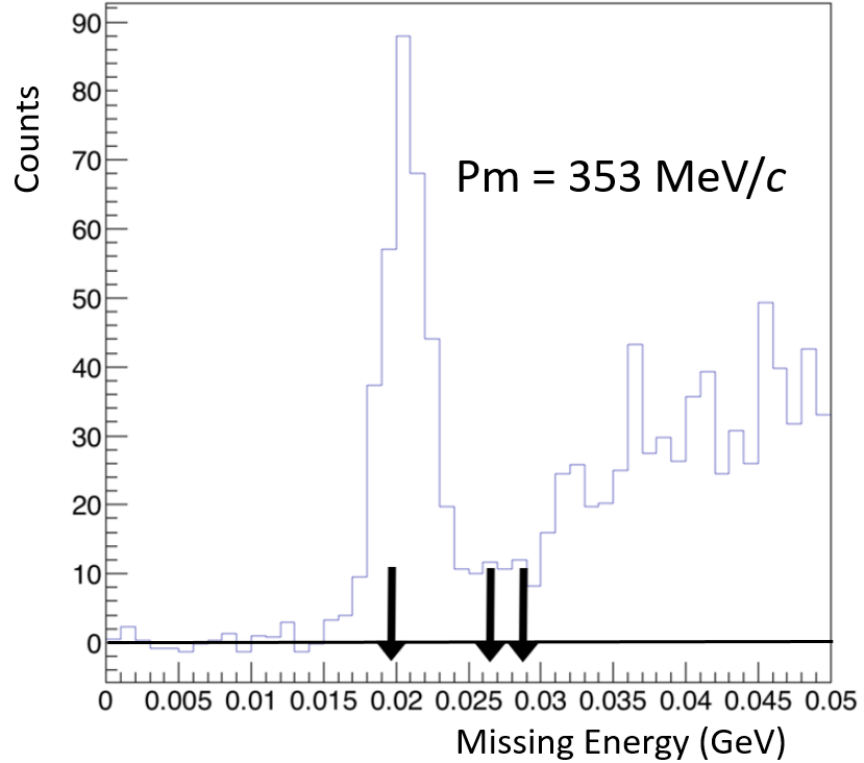}
		\label{fig353}
		\includegraphics[scale=0.4]{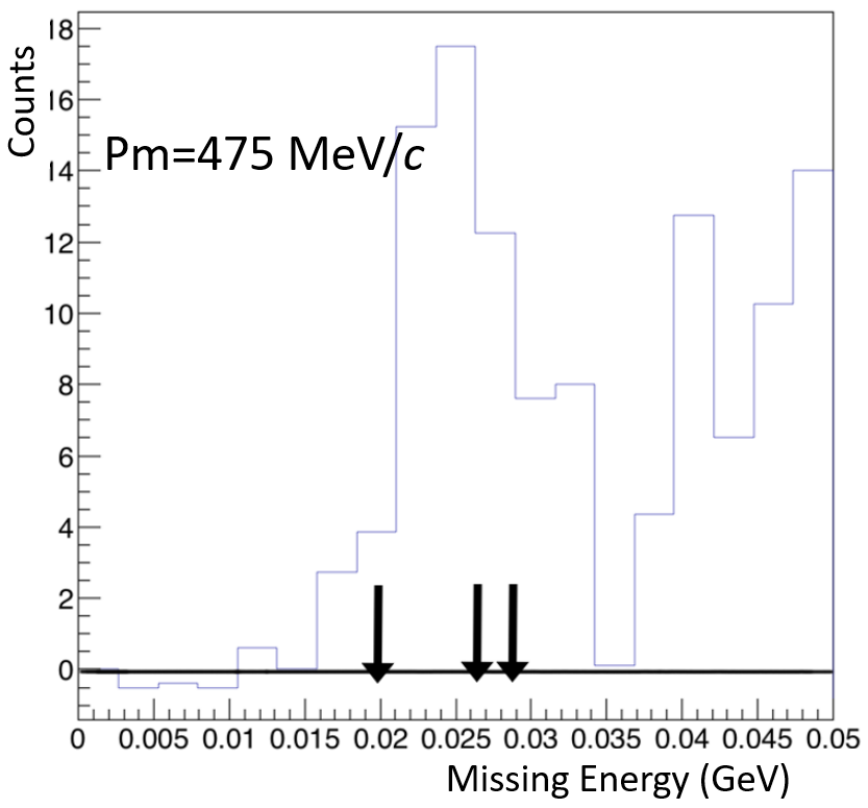}
		\label{fig475}
		\includegraphics[scale=0.4]{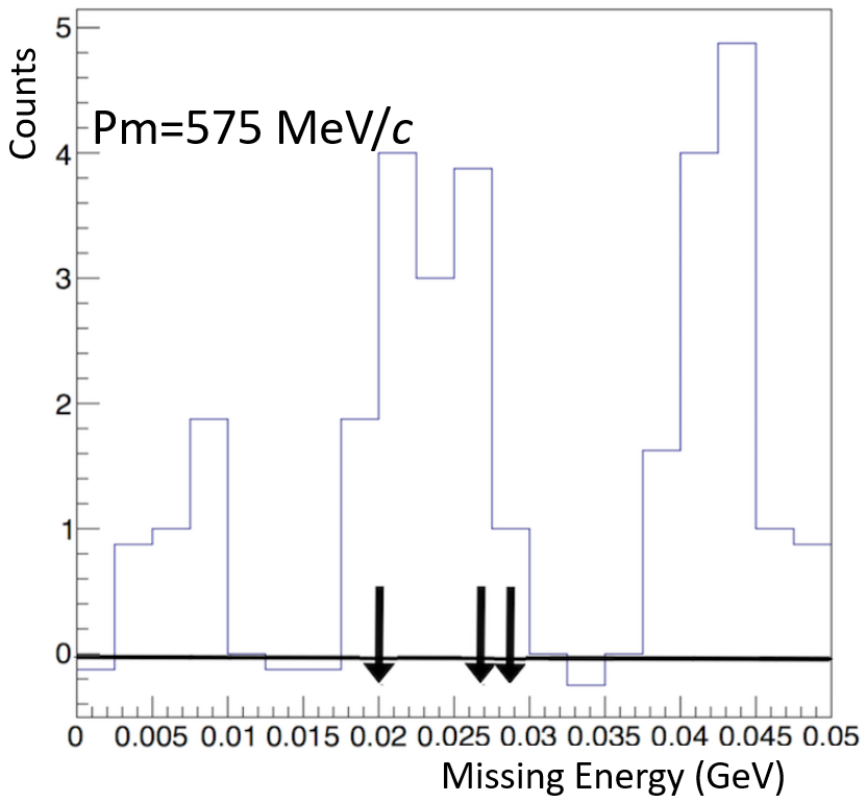}
		\caption{From top to bottom: Missing energy region up to 50 MeV of excitation in $^4He(e,e'p)X$ for
			$p_m$=153, 352, 475 and 575 MeV/$c$, respectively. The three arrows point to the expected locations of the thresholds of the hadronic
			states X=(t), X=(n,d) and X=(p,n,n).}
		\label{fig575}
	\end{figure}

Portions of the missing energy spectrum in the triton/3N energy range are shown in figures~ \ref{fig575}. We see a change in the distribution of events as a function of missing momentum going from  153 MeV/$c$ to 575~MeV/$c$. At low missing momenta the triton peak is centered at the expected value of 19.8 MeV. At higher missing momenta, the events are higher in missing energy by a few MeVs. From left to right, the three arrows in each figure point to the expected locations of the thresholds of the hadronic states X=(t), X=(n,d) and X=(p,n,n), respectively. 

 An interesting question is the impact of three-nucleon forces, $V_{ijk}$, at high $p_m$. $V_{ijk}$ are known to increase the binding energy of nuclei~\cite{3NN} so they would be natural actors in the formation of bound tritons or closely bound three nucleon groups among the outgoing
hadronic channels, $X$, at high missing momentum. The principal sources of data to help refine models of possible three-nucleon interactions are binding energies of ground and excited states of $A<8$ nuclei and point proton charge distributions ~\cite{3NN}. However, these data
are not extensive enough to select unambiguously  a particular set of parameters or models for $V_{ijk}$  and other observables are needed as discussed in ~\cite{3NN}~\cite{VMC}.
 
 More extensive and detailed data in the three nucleon triton mass region and the existence of microscopic calculations for these nuclei opens the possibility of exploiting the shapes of the missing energy spectra in $A(e,e'p)X$ reactions as additional observables for developing models of three-nucleon interactions.

\begin{acknowledgments}
		Special thanks to Silviu Covrig for providing the CFD calculations as a possibility to understanding the target vertex spectra for the SRC Target. Special thanks to Or Hen for valuable discussions and inputs on the paper. The research presented in this paper is partially supported by the U.S. National Science Foundation grants PHY 09-69380 and PHY 16-15067. This work was supported by the U.S. Department of Energy contract DE-AC05-06OR23177 under which Jefferson Science Associates operates the Thomas Jefferson National Accelerator Facility.	
	\end{acknowledgments}

\section{Tabulated results \label{cross}}

Table~\ref{tab:uncertainties} summarizes the location of the parameters involved in the cross section in this document or their uncertainties. 
Experimental differential cross sections for $^4He(e,e'p)X$;
 in $nb/sr^2/MeV$, are summarized in table \ref{tab:xsect} for the four different spectrometer settings 
Tables \ref{tab:madridEMA} and \ref{tab:madridfull} summarize the Madrid EMA and full calculations respectively in the momentum range from 12.5 to 637.5 MeV/$c$. 

\begin{table}[h!]
\begin{tabular}{|c|c|}
\hline
parameter & location or values\\
\hline
RSC & sect IIIC and sect IIID, 4b\\
$\Delta \Omega_e$ & 4.8msr $\pm 2.7\%$, Fig. 4\\
$\Delta \Omega_p$ & 4.8msr $\pm 2.7\%$, Fig. 4\\
$\Delta E_e$ & sect IIID, 4d, $\pm 10\%$, Fig.8\\
$N_e$   & sect IIID, 4e\\
$N_{tgt}$ & sect IIID, 4f and Table II\\
Eff  & sect IIID, 4g\\
$f(p_m)$  & equation 3, discussed in  sect IIIB, sect IIID 4g.1\\
LT(daq)  & sect IIID, 4g.2, depends on proton $p_m$ setting\\
LT(el)   & sect IIID, 4g.3, Table III\\
Tri  & sect IIID, 4g.4, Table III\\
WC   & sect IIID, 4g.5, Table III\\
Tra  & sect IIID, 4g.6, Table III\\
\hline
\end{tabular}
\caption{Parameters in the cross section calculation and their uncertainties.}
\label{tab:uncertainties}
\end{table}

\begin{sidewaystable} \label{TableVI}	
	\begin{tabular}{c|c|c|c|c}
		\hline
		$p_m$ &  153 & 353 & 466 & 632  \\
		(MeV/$c$)    &$\theta_p=47^\circ$ &$\theta_p=38.5^\circ$
		&$\theta_p=33.5^\circ$ & $\theta_p=29^\circ$\\
		&             &     $\times 10^{-4}$  &   $\times 10^{-4}$   &  $\times
		10^{-5}$ \\ 
		\hline
		25    &$3.38\pm0.40\pm0.48\pm0.62$&  &  &  \\
		75    &$1.13\pm0.13\pm0.16\pm0.20$&  &  &  \\
		125   &$(3.13\pm0.36\pm0.44\pm0.57)\times 10^{-1}$&  &  &  \\
		175   &$(7.18\pm0.83\pm1.02\pm1.31)\times 10^{-2}$&  &  &  \\
		225  &$(1.44\pm0.17\pm0.20\pm0.26)\times
		10^{-2}$&$44.0\pm1.4\pm6.2\pm6.4$  &  &  \\
		275  &$(3.06\pm0.48\pm0.43\pm0.65)\times
		10^{-3}$&$12.7\pm1.3\pm1.8\pm2.2$&   &   \\
		325  &    &$6.11\pm0.14\pm0.86\pm0.88$ &  &  \\
		375  &    &$3.57\pm0.80\pm0.50\pm0.95$    &    &    \\
		425  &    &$1.44\pm0.57\pm\pm0.20\pm0.61$   
		&$(6.59\pm2.62\pm0.93\pm2.78)$    &    \\
		475  &    &    &$(3.22\pm0.83\pm0.46\pm0.95)$    &    \\
		525  &    &    &$(1.68\pm0.42\pm0.24\pm0.48)$    &    \\
		575  &    &    &$(0.91\pm0.42\pm0.13\pm0.44)$    &    \\
		632  &    &    &    &$(3.7\pm2.27\pm0.52\pm2.33)$    \\
		\hline
	\end{tabular}
	\caption{Experimental differential cross sections, $\frac{d
			\sigma^5}{d\Omega_p d\Omega_e dE_e}$,
		for $^4He(e,e'p)X$; where  $X$ $=$ $^3H$ or $3N$,  from E08009, for
		different
		kinematical settings given by the proton spectrometer central angle. Statistical
		and normalization uncertainties are the first
		uncertainty entry, systematic uncertainties in selecting the size of the $\Delta E$
		bin of 10\% and an estimated 10\% from the
		missing momentum acceptance factor, described in sections IIIB and IIID(d) are the second
		uncertainty entry,
		total uncertainty is the third entry. Units are $nb/sr^2/MeV$.}
	\label{tab:xsect}
\end{sidewaystable}
\begin{sidewaystable}
		\begin{tabular}{c|c|c|c|c}
			\hline
			$p_m$ &  ~~$\overline{p_e}$~~~~~~~$\overline{\theta_e}$~~~~$\overline{\theta_p}$~~~$\overline{p_{m}}$~ & ~~$\overline{p_e}$~~~~~~~$\overline{\theta_e}$~~~~$\overline{\theta_p}$~~~$\overline{p_{m}}$~&~~$\overline{p_e}$~~~~~~~$\overline{\theta_e}$~~~~$\overline{\theta_p}$~~~$\overline{p_{m}}$~ &~~$\overline{p_e}$~~~~~~~$\overline{\theta_e}$~~~~$\overline{\theta_p}$~~~$\overline{p_{m}}$~ \\
			({\tiny{MeV/$c$}}) &   ({\tiny{GeV/$c$}}) ~ rad ~ rad ~ rad~  &   ({\tiny{GeV/$c$}}) ~ rad ~ rad ~ rad~   &   ({\tiny{GeV/$c$}}) ~ rad ~ rad ~ rad~   & ({\tiny{GeV/$c$}}) ~ rad ~ rad ~ rad~\\
			\hline
			25    &$3.482~~0.338~~0.807~~40.1$  &  &  \\
			75    &$3.495~~0.345~~0.817~~81.0$&  &  &  \\
			125   &$3.503~~0.351~~0.822~~126.6$&  &  &  \\
			175   &$3.527~~0.353~~0.822~~175.2$&  &  &  \\
			225  &$3.557~~0.355~~0.822~~224.8$&$3.454~~0.350~~0.675~~226.8$  &  &  \\
			275  &$3.591~~0.356~~0.821~~274.3$&$3.488~~0.353~~0.673~~275.6$&   &   \\
			325  &    &$3.525~~0.355~~0.672~~324.9$ &  &  \\
			375  &    &$3.565~~0.355~~0.672~~374.8$    &    &    \\
			425  &    &$3.603~~0.356~~0.672~~424.6$    &$3.496~~0.355~~0.585~~425.3$    &    \\
			475  &    &                                &$3.542~~0.355~~0.585~~474.9$    &    \\
			525  &    &                                &$3.586~~0.356~~0.585~~524.5$    &    \\
			575  &    &                                &$3.619~~0.359~~0.584~~573.8$    &    \\
			632  &    &    &    &$3.604~~0.360~~0.505~~676.6$    \\
			\hline
		\end{tabular}
	\label{tab:pmaverage}
		\caption{Average values from the GEANT simulation of some kinematic variables
			in each of the 50~MeV wide missing momentum bins used in the analysis. The bins are identified by the algebraic midpoints, for example, $p_m=25$ is for the missing momentum bin from 0 to 50~MeV/$c$. The entries in each line are $<p_e>$ in $GeV/c$, $<\theta_e>$ in radians,
			$<\theta_p>$ in radians and $<p_m>$ in $MeV/c$.  }
\end{sidewaystable}

\begin{table*}[htp]
	\begin{tabular}{c|c|c|c|c}
		\hline
		$p_m$ &  153 & 353 & 466 & 632  \\
		(MeV/c)    &$\theta_p=47^\circ$ &$\theta_p=38.5^\circ$ &$\theta_p=33.5^\circ$ &
		$\theta_p=29^\circ$\\
		\hline
		12.5 &2.2059 & & & \\
		37.5 &1.8287 & & & \\
		62.5 &1.3139 & & & \\
		87.5 &8.516e-01 & & & \\
		112.5 &5.070e-01 & & & \\
		137.5 &2.699e-01 & & & \\
		162.5 &1.311e-01 & & & \\
		187.5 &5.987e-02 & & & \\
		212.5 &2.583e-02 &1.918e-02 & & \\
		237.5 &1.044e-02 &6.724e-03 & & \\
		262.5 &3.951e-03 &2.209e-03 & & \\
		287.5 &1.370e-03 &6.686e-04 & & \\
		312.5 &4.901e-04 &3.578e-04 & & \\
		337.5 &1.858e-04 &3.095e-04 & & \\
		362.5 &9.309e-05 &2.687e-04 & & \\
		387.5 &5.639e-05 &2.077e-04 & & \\
		412.5 & &1.419e-04 &5.283e-04 & \\
		437.5 & &8.366e-05 &3.402e-04 & \\
		462.5 & &4.808e-05 &2.225e-04 & \\
		487.5 & &2.739e-05 &1.262e-04 &2.206e-04 \\
		512.5 & &1.542e-05 &6.542e-05 &1.491e-04 \\
		537.5 & &9.478e-06 &2.980e-05 &8.585e-05 \\
		562.5 & & &1.289e-05 &4.400e-05 \\
		587.5 & & &5.077e-06 &1.977e-05 \\
		612.5 & & &2.008e-06 &7.741e-06 \\
		637.5 & & &8.357e-07 &2.834e-06 \\
		\hline
	\end{tabular}
	\caption{Madrid full theoretical cross sections integrated over the
		experimental acceptances
		for $^4He(e,e'p)^3H$ for E08009, for different
		kinematical settings given by the proton spectrometer central angle.
		Units are $nb/sr^2/MeV$.}
	\label{tab:madridfull}
\end{table*}

\begin{table*}[htp]
	\begin{tabular}{c|c|c|c|c}
		\hline
		\textbf{$p_m$}  &  153 & 353 & 466 & 632  \\
		(MeV/$c$)    &$\theta_p=47^\circ$ &$\theta_p=38.5^\circ$ &$\theta_p=33.5^\circ$ & $\theta_p=29^\circ$\\
		\hline
		37.5 &2.681 & & & \\
		62.5 &1.916 & & & \\
		87.5 &1.235 & & & \\
		112.5 &7.297e-01 & & & \\
		137.5 &3.839e-01 & & & \\
		162.5 &1.834e-01 & & & \\
		187.5 &8.159e-02 &9.031e-02 & & \\
		212.5 &3.382e-02 &3.628e-02 & & \\
		237.5 &1.282e-02 &1.295e-02 & & \\
		262.5 &4.433e-03 &3.933e-03 & & \\
		287.5 &1.362e-03 &9.986e-04 & & \\
		312.5 &4.312e-04 &3.423e-04 & & \\
		337.5 &1.705e-04 &2.643e-04 & & \\
		362.5 &1.130e-04 &2.487e-04 & & \\
		387.5 &8.817e-05 &2.083e-04 & & \\
		412.5 & &1.547e-04 &4.550e-04 & \\
		437.5 & &9.853e-05 &3.082e-04 & \\
		462.5 & &6.482e-05 &2.064e-04 & \\
		487.5 & &4.261e-05 &1.206e-04 &1.778e-04 \\
		512.5 & & &6.435e-05 &1.215e-04 \\
		537.5 & & &3.036e-05 &7.084e-05 \\
		562.5 & & &1.360e-05 &3.702e-05 \\
		587.5 & & &5.527e-06 &1.717e-05 \\
		612.5 & & &2.251e-06 &7.010e-06 \\
		637.5 & & &9.483e-07 &2.695e-06 \\
		\hline
	\end{tabular}
	\caption{Madrid EMA theoretical cross sections integrated over the experimental acceptances for $^4He(e,e'p)^3H$ for E08009, for different kinematical settings given by the proton spectrometer central angle. Units are $nb/sr^2/MeV$.}
	\label{tab:madridEMA}	
\end{table*}

\bibliography{e08009}

\end{document}